\documentclass[journal,letterpaper,final,onecolumn,romanappendices]{IEEEtran}
\IEEEoverridecommandlockouts

\ifCLASSINFOpdf
\else
\fi
\usepackage[tight,footnotesize]{subfigure}

\usepackage{amsmath,amssymb,amsbsy,amsthm}
\usepackage{bm}
\usepackage{fixmath}
\usepackage{cite}
\usepackage{hyperref}
\usepackage{graphicx}

\newcommand{\BA}{\B{A}}

\newcommand{\BQ}{\B{Q}}

\newcommand{\BS}{\B{S}}
\newcommand{\BT}{\B{T}}
\newcommand{\BU}{\B{U}}

\DeclareMathAlphabet{\mathbit}{OML}{cmr}{bx}{it}
\DeclareMathAlphabet{\mathgoth}{U}{ygoth}{m}{n}
\DeclareMathAlphabet{\mathfrak}{U}{yfrak}{m}{n}
\DeclareMathAlphabet{\mathswab}{U}{yswab}{m}{n}

\newcommand{\B}[1]{\mathbf{#1}}

\newcounter{thm_counter}
\newcounter{rmk_counter}
\newtheorem{mypro}[thm_counter]{Proposition}
\newtheorem{mythm}[thm_counter]{Theorem}

\newtheorem{mycor}[thm_counter]{Corollary}
\newtheorem{mylem}[thm_counter]{Lemma}

\newtheorem{myrmk}[rmk_counter]{Remark}

\DeclareMathOperator{\Prob}{\mathsf{Pr}}
\DeclareMathOperator{\Exp}{\mathsf{E}}
\newcommand{\intd}{{\,\operatorname{d}}}
\newcommand{\diffd}{{\operatorname{d}}}
\newcommand{\e}{{\operatorname{e}}}

\newcommand{\dB}{{\operatorname{dB}}}

\DeclareMathOperator{\expint}{Ei} 

\begin{document}
\title{A Rate-Splitting Approach to Fading Channels\\with Imperfect Channel-State Information}

\author{
\IEEEauthorblockN{
Adriano~Pastore,~\IEEEmembership{Student Member,~IEEE},
Tobias~Koch,~\IEEEmembership{Member,~IEEE},\\
and Javier~Rodr\'{i}guez~Fonollosa,~\IEEEmembership{Senior Member,~IEEE}
}
\thanks{A.~Pastore and J.~R.~Fonollosa have been supported by the Ministerio de Econom\'ia y Competitividad of Spain (TEC2010-19171 and CONSOLIDER-INGENIO 2010 CSD2008-00010 COMONSENS) and Generalitat de Catalunya (2009SGR-1236). T.~Koch has been supported by the European Community's Seventh's Framework Programme (FP7/2007-2013) under grant agreement No.~333680 and by
the Ministerio de Econom\'ia y Competitividad of Spain (TEC2009-14504-C02-01, TEC2012-38800-C03-01,  and CONSOLIDER-INGENIO 2010 CSD2008-00010 COMONSENS). The material in this paper was presented in part at the IEEE 27th Convention of Electrical and Electronics Engineers in Israel, Eilat, Israel, November 14--17, 2012.}
\thanks{A.~Pastore and J.~R.~Fonollosa are with the Department of Signal Theory and Communications, Universitat Polit\`{e}cnica de Catalunya, Jordi Girona 1--3, Building D5, 08034 Barcelona, Spain (email: \{adriano.pastore,javier.fonollosa\}@upc.edu).}
\thanks{T.~Koch is with the Signal Theory and Communications Department, Universidad Carlos III de Madrid, 28911 Legan\'es, Spain (email: koch@tsc.uc3m.es).}
}

\sloppy

\maketitle
\IEEEpeerreviewmaketitle


\begin{abstract}
As shown by M\'edard, the capacity of fading channels with imperfect channel-state information (CSI) can be lower-bounded by assuming a Gaussian channel input $X$ with power $P$ and by upper-bounding the conditional entropy $h(X|Y,\hat{H})$ by the entropy of a Gaussian random variable with variance equal to the linear minimum mean-square error in estimating $X$ from $(Y,\hat{H})$. We demonstrate that, using a rate-splitting approach, this lower bound can be sharpened: by expressing the Gaussian input $X$ as the sum of two independent Gaussian variables $X_1$ and $X_2$ and by applying M\'edard's lower bound first to bound the mutual information between $X_1$ and $Y$ while treating $X_2$ as noise, and by applying it a second time to the mutual information between $X_2$ and $Y$ while assuming $X_1$ to be known, we obtain a capacity lower bound that is strictly larger than M\'edard's lower bound. We then generalize this approach to an arbitrary number $L$ of layers, where $X$ is expressed as the sum of $L$ independent Gaussian random variables of respective variances $P_\ell$, $\ell = 1,\dotsc,L$ summing up to $P$. Among all such rate-splitting bounds, we determine the supremum over power allocations $P_\ell$ and total number of layers $L$. This supremum is achieved for $L \rightarrow \infty$ and gives rise to an analytically expressible capacity lower bound. For Gaussian fading, this novel bound is shown to converge to the Gaussian-input mutual information as the signal-to-noise ratio (SNR) grows, provided that the variance of the channel estimation error $H-\hat{H}$ tends to zero as the SNR tends to infinity.
\end{abstract}

\section{Introduction and Channel Model} \label{introduction}

We consider a single-antenna memoryless fading channel with imperfect channel-state information (CSI), whose time-$k$ channel output $Y[k]$ corresponding to a time-$k$ channel input $X[k] = x\in\mathbb{C}$ (where $\mathbb{C}$ denotes the set of complex numbers) is given by
\begin{align}   \label{system_equation}
	Y[k] = \bigl(\hat{H}[k]+\tilde{H}[k]\bigr) \, x + Z[k], \quad k\in\mathbb{Z}
\end{align}
(with $\mathbb{Z}$ denoting the set of integers). Here, the noise $\{Z[k]\}_{k\in\mathbb{Z}}$ is a sequence of independent and identically distributed (i.i.d.), zero-mean, circularly-symmetric, complex Gaussian random variables with variance $\Exp\bigl[|Z[k]|^2\bigr] = N_0$. The fading pair $\bigl\{(\hat{H}[k],\tilde{H}[k])\bigr\}_{k\in\mathbb{Z}}$ is an arbitrary sequence of i.i.d.\ complex-valued random variables whose means and variances satisfy the following conditions: 
\begin{itemize}
\item $\hat{H}[k]$ has mean $\mu$ and variance $\hat{V}$;
\item conditioned on $\hat{H}[k]=\hat{h}$, the random variable $\tilde{H}[k]$ has zero mean and variance $\tilde{V}(\hat{h})$, i.e.,
\begin{subequations}
\begin{IEEEeqnarray}{rCl}
	\Exp\bigl[\tilde{H}[k] \bigm| \hat{H}[k] = \hat{h} \bigr]
	& = & 0   \label{conditional_unbiasedness} \\
	\Exp\bigl[ |\tilde{H}[k]|^2 \bigm| \hat{H}[k] = \hat{h} \bigr]
	&\triangleq & \tilde{V}(\hat{h}).
\end{IEEEeqnarray}
\end{subequations}
\end{itemize}
We assume that the joint sequence $\bigl\{(\hat{H}[k],\tilde{H}[k])\bigr\}_{k\in\mathbb{Z}}$, the noise sequence $\{Z[k]\}_{k\in\mathbb{Z}}$ and the input sequence $\{X[k]\}_{k\in\mathbb{Z}}$ are independent. We further assume that the receiver is cognizant of the realization of $\{\hat{H}[k]\}_{k\in\mathbb{Z}}$, but the transmitter is only cognizant of its distribution. We finally assume that both the transmitter and receiver are cognizant of the distributions of $\{\tilde{H}[k]\}_{k\in\mathbb{Z}}$ and $\{Z[k]\}_{k\in\mathbb{Z}}$ but not of their realizations.

The $\hat{H}[k]$ can be viewed as an estimate of the fading coefficient
\begin{align}
	H[k] \triangleq \hat{H}[k]+\tilde{H}[k].
\end{align}
Accordingly, $\tilde{H}[k]$ can be viewed as the channel estimation error. From this perspective, the condition \eqref{conditional_unbiasedness} is, for example, satisfied when $\hat{H}[k]$ is the minimum mean-square error (MMSE) estimate of $H[k]$ from some receiver side information independent of the input $X[k]$. When $\tilde{H}[k] = 0$ almost surely, we shall say that the receiver has perfect CSI.

The capacity of the above channel \eqref{system_equation} under the average-power constraint $P$ on the channel inputs is given by \cite{BiProSh98}
\begin{align}
\label{capacity}
C(P) = \sup I(X;Y|\hat{H})
\end{align}
where the supremum is over all distributions of $X$ satisfying $\Exp[|X|^2]\leq P$. Here and throughout the paper we omit the time indices $k$ wherever they are immaterial. Since \eqref{capacity} is difficult to evaluate, even if $\hat{H}$ and $\tilde{H}$ are Gaussian, it is common to assess $C(P)$ using upper and lower bounds. A widely-used lower bound on $C(P)$ is due to M\'edard~\cite{Me00}:
\begin{align} \label{Medard_lower_bound}
C(P) \geq  \Exp\left[\log\left(1 + \frac{|\hat{H}|^2 P}{\tilde{V}(\hat{H}) P + N_0}\right)\right] \triangleq R_{\textnormal{M}}(P).
\end{align}
Here and throughout this paper, $\log(\cdot)$ denotes the natural logarithm function. Consequently, all rates specified in this paper are in nats per channel use. The lower bound \eqref{Medard_lower_bound} follows from \eqref{capacity} by choosing the input $X_\textnormal{G}$ to be zero-mean, variance-$P$, circularly-symmetric, complex Gaussian\footnote{The subscript `G' in $X_\textnormal{G}$ indicates a Gaussian distribution.} and by upper-bounding the differential entropy of $X_\textnormal{G}$ conditioned on $Y$ and $\hat{H}$ as
\begin{align}   \label{h_upperbound}
	h(X_\textnormal{G}|Y,\hat{H}) & = h(X_\textnormal{G}-\alpha Y|Y,\hat{H}) \nonumber\\
	& \leq  h(X_\textnormal{G}-\alpha Y|\hat{H}) \nonumber\\
	& \leq \Exp\left[\log\left(\pi e \Exp\bigl[|X_\textnormal{G}-\alpha Y|^2\bigm|\hat{H}\bigr]\right)\right]
\end{align}
for any $\alpha\in\mathbb{C}$. Here the first inequality follows because conditioning cannot increase entropy, and the subsequent inequality follows because the Gaussian distribution maximizes differential entropy for a given second moment \cite[Theorem~9.6.5]{coverthomas91}. By expressing the mutual information $I(X_\textnormal{G};Y|\hat{H})$ as
\begin{align}   \label{MI_expansion}
	I(X_\textnormal{G};Y|\hat{H}) = h(X_\textnormal{G}) - h(X_\textnormal{G}|Y,\hat{H})
\end{align}
and by choosing $\alpha$ in \eqref{h_upperbound} so that $\alpha Y$ is the linear MMSE estimate of $X_\textnormal{G}$, the lower bound \eqref{Medard_lower_bound} follows.

When the receiver has perfect CSI so that $\Exp[\tilde{V}(\hat{H})]=0$, the lower bound $R_{\textnormal{M}}(P)$ is equal to the channel capacity
\begin{equation}   \label{coherent_capacity}
	C_\textnormal{coh}(P) = \Exp\left[\log\left(1+\frac{|H|^2 P}{N_0}\right)\right].
\end{equation}
Consequently, for perfect CSI the lower bound \eqref{Medard_lower_bound} is tight.  In contrast, when the receiver has imperfect CSI and the distributions of $\tilde{V}(\hat{H})$ and $\hat{H}$ do not depend on $P$, the lower bound \eqref{Medard_lower_bound} is loose. In fact, in this case $R_{\textnormal{M}}(P)$ is bounded in $P$, whereas the capacity $C(P)$ is known to be unbounded. For instance, if the conditional entropy of $\tilde{H}$ given $\hat{H}$ is finite, then the capacity has a double-logarithmic growth in $P$ \cite{LaMo03}.\footnote{This result can be generalized to show that if $\Exp[\log|\hat{H}+\tilde{H}|^2]>-\infty$ holds, then the capacity grows at least double-logarithmically with $P$.} 

This boundedness of $R_{\textnormal{M}}(P)$ is not due to the inequalities in \eqref{h_upperbound} being loose, but is a consequence of choosing a Gaussian channel input. Indeed, if $h(\tilde{H}|\hat{H})$ is finite, then a Gaussian input $X_\textnormal{G}$ achieves \cite[Proposition~6.3.1 and Lemma~6.2.1]{LaSh02} (see also \cite[Lemma 4.5]{LaMo03})
\begin{equation}
	\varlimsup_{P\to\infty} I(X_{\textnormal{G}};Y|\hat{H}) \leq \gamma + \log\bigl(\pi e \Exp\bigl[|\hat{H}+\tilde{H}|^2\bigr]\bigr) - h(\tilde{H}|\hat{H})
\end{equation}
where $\gamma\approx 0.577$ denotes Euler's constant and where $\varlimsup$ denotes the \emph{limit superior}. Nevertheless, even if we restrict ourselves to Gaussian inputs, the lower bound
\begin{equation}   \label{Medard_lower_bound_2}
	I(X_{\textnormal{G}};Y|\hat{H}) \geq R_{\textnormal{M}}(P)
\end{equation}
is not tight. As we shall see, by using a rate-splitting and successive-decoding approach, this lower bound \eqref{Medard_lower_bound_2} can be sharpened: we show in Section~\ref{sec:two_layers} that, by expressing the Gaussian input $X_{\textnormal{G}}$ as the sum of two independent Gaussian random variables $X_1$ and $X_2$, and by first applying the bounding technique sketched in \eqref{h_upperbound}--\eqref{MI_expansion} to $I(X_1;Y|\hat{H})$ (thus treating $H X_2$ as noise) and then using the same bounding technique to lower-bound $I(X_2;Y|\hat{H},X_1)$, we obtain a lower bound on the Gaussian-input mutual information (and thus also on the capacity) that is strictly larger than the conventional bound $R_{\textnormal{M}}(P)$.

In Section~\ref{sec:L-layering}, we extend this approach by expressing $X_\textnormal{G}$ as the sum of $L \geq 2$ independent Gaussian random variables $X_{\ell}$, $\ell=1,\ldots,L$ and by applying the bounding technique from \eqref{h_upperbound}--\eqref{MI_expansion} first to $I(X_1;Y|\hat{H})$, then to $I(X_2;Y|\hat{H},X_1)$, and so on. We show that the so obtained lower bound is strictly increasing in $L$ (provided that we optimize the sum of bounds over the powers $P_\ell = \Exp[|X_{\ell}|^2]$, $\ell=1,\ldots,L$), and we determine its limit as $L$ tends to infinity. The so-obtained lower bound permits an analytic expression. In the remainder of this paper, we shall refer to the index $\ell$ as a \emph{layer} and to $L$ as the \emph{number of layers}.

In Section~\ref{sec:asymptotically_optimal_csi}, we show that when, conditioned on $\hat{H}$, the estimation error $\tilde{H}$ is Gaussian, and when its variance (averaged over $\hat{H}$) tends to zero as the SNR tends to infinity, the new lower bound tends to the Gaussian-input mutual information $I(X_\textnormal{G};Y|\hat{H})$ as the SNR tends to infinity. For non-Gaussian fading, we show that, at high SNR, the difference between $I(X_\textnormal{G};Y|\hat{H})$ and our lower bound is upper-bounded by the difference of the logarithms of the variance of $\tilde{H}$ and of its entropy power.

In Section~\ref{sec:mismatch} we discuss the connection of our results with similar results obtained in the mismatched-decoding literature, and in Section~\ref{sec:summary} we conclude the paper with a summary and discussion. In Appendices~\ref{app:III} and \ref{app:IV} we provide the proofs of the main results from Sections~\ref{sec:L-layering} and \ref{sec:asymptotically_optimal_csi}, respectively.

\section{Rate-Splitting with Two Layers}   \label{sec:two_layers}

For future reference, we state M\'edard's lower bound \eqref{Medard_lower_bound} in a slightly more general form in the following proposition.

\begin{mypro}
\label{pro:Medard_bound}
Let $S$ be a zero-mean, circularly-symmetric, complex Gaussian random variable of variance $P$. Let $A$ and $B$ be complex-valued random variables of finite second moments, and let $C$ be an arbitrary random variable. Assume that $S$ is independent of $(A,C)$, and that, conditioned on $(A,C)$, the variables $S$ and $B$ are uncorrelated. Then
\begin{align}   \label{Medard_bound}
	I(S;AS+B|A,C)
	\geq \Exp\left[\log\left(1 + \frac{|A|^2 P}{V_{B}(A,C)} \right) \right]
\end{align}
where $V_B(a,c)$ denotes the conditional variance of $B$ conditioned on $(A,C) = (a,c)$.
\end{mypro}

\begin{IEEEproof}
See Appendix~\ref{app:pro:proof:Medard_bound}.
\end{IEEEproof}

Using Proposition~\ref{pro:Medard_bound}, we show that, for imperfect CSI and $\Exp[|\hat{H}|^2]>0$, rate splitting with two layers strictly improves the lower bound \eqref{Medard_lower_bound_2}. Indeed, let $X_1$ and $X_2$ be independent, zero-mean, circularly-symmetric, complex Gaussian random variables with respective variances $P_1$ and $P_2$ (satisfying $P_1+P_2=P$) such that $X_\textnormal{G}=X_1+X_2$. By the chain rule for mutual information, we obtain
\begin{align}   \label{chain_rule_for_two_layers}
	I(X_\textnormal{G};Y|\hat{H}) & = I(X_1,X_2;Y|\hat{H}) \nonumber\\
	& = I(X_1;Y|\hat{H}) + I(X_2;Y|\hat{H},X_1).
\end{align}
By replacing the random variables $A$, $B$, $C$, and $S$ in Proposition~\ref{pro:Medard_bound} with
\begin{equation*}
	A \gets \hat{H}, \quad B\gets \hat{H}X_2 + \tilde{H}X + Z, \quad C \gets 0, \quad S\gets X_1
\end{equation*}
and by noting that these random variables satisfy the proposition's conditions, it follows that the first term on the right-hand side (RHS) of \eqref{chain_rule_for_two_layers} is lower-bounded as
\begin{align}   \label{R1}   
	I(X_1;Y|\hat{H})
	\geq \Exp\left[\log\left( 1 + \frac{|\hat{H}|^2 P_1}{\tilde{V}(\hat{H})P_1 + \bigl(|\hat{H}|^2 + \tilde{V}(\hat{H})\bigr)P_2 + N_0} \right)\right]
	\triangleq R_1(P_1,P_2).
\end{align}
Similarly, by replacing $A$, $B$, $C$, and $S$ in Proposition~\ref{pro:Medard_bound} with
\begin{equation*}
	A \gets \hat{H}, \quad B \gets \hat{H}X_1 + \tilde{H}X + Z, \quad C \gets X_1, \quad S \gets X_2
\end{equation*}
and by noting that these random variables satisfy the proposition's condition, we obtain for the second term on the RHS of \eqref{chain_rule_for_two_layers}
\begin{align}   \label{R2}   
	I(X_2;Y|\hat{H},X_1)
	\geq \Exp\left[\log\left( 1 + \frac{|\hat{H}|^2 P_2}{\tilde{V}(\hat{H})(|X_1|^2 + P_2) + N_0} \right)\right]
	\triangleq R_2(P_1,P_2).
\end{align}
Since for every $\alpha>0$, the function $x\mapsto \log(1+\alpha/x)$ is strictly convex in $x>0$, it follows from
Jensen's inequality that the RHS of \eqref{R2} is lower-bounded as
\begin{align}   
	\Exp\left[\log\left( 1 + \frac{|\hat{H}|^2 P_2}{\tilde{V}(\hat{H})(|X_1|^2 + P_2) + N_0} \right)\right]
	\geq \Exp\left[\log\left( 1 + \frac{|\hat{H}|^2 P_2}{\tilde{V}(\hat{H})(P_1+P_2)+N_0}\right)\right] \label{R2'}
\end{align}
with the inequality being strict except in the trivial cases where $P_1=0$, $P_2=0$, or if, with probability one, at least one of $|\hat{H}|$ and $\tilde{V}(\hat{H})$ is zero.\footnote{We shall write this as $\Prob\bigl\{\hat{H} \cdot \tilde{V}(\hat{H}) = 0\bigr\} = 1$. For example, this occurs when the receiver has perfect CSI, in which case $\tilde{V}(\hat{H}) = 0$ almost surely.} Thus, combining \eqref{chain_rule_for_two_layers}--\eqref{R2'}, we obtain
\begin{align}
	R_1(P_1,P_2) + R_2(P_1,P_2)
	\geq \Exp\left[\log\left( 1 + \frac{|\hat{H}|^2 P}{\tilde{V}(\hat{H})P + N_0} \right)\right]
\end{align}
demonstrating that, when the receiver has imperfect CSI, rate splitting with two layers strictly improves the lower bound \eqref{Medard_lower_bound} (except in trivial cases). 

\begin{figure}[h]
\begin{center}
	\includegraphics{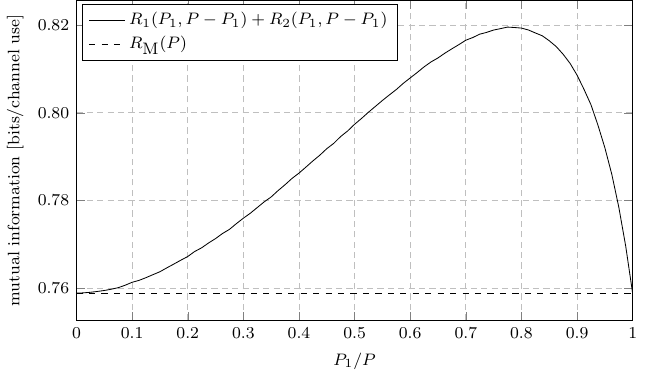}
	\caption{Comparison of the two-layer lower bound $R_1(P_1,P-P_1)+R_2(P_1,P-P_1)$ (continuous line) with M\'edard's lower bound $R_{\textnormal{M}}(P)$ (dashed line) as a function of the power fraction $P_1/P$ assigned to the first layer.}
	\label{fig:two_layers}
\end{center}
\end{figure}

Figure~\ref{fig:two_layers} compares the two-layer bound $R_1(P_1,P_2) + R_2(P_1,P_2)$ with $R_{\textnormal{M}}(P)$ (dashed line) as a function of $P_1/P$, for $\hat{H}$ and $\tilde{H}$ being mutually independent and circularly-symmetric Gaussian with parameters $\mu = 0$, $\hat{V} = \frac{1}{2}$,  $\tilde{V}(\hat{h}) = \frac{1}{2}$ for $\hat{h}\in\mathbb{C}$, $P = 10$, and $N_0 = 1$. The figure confirms our observation that, when the receiver has imperfect CSI and $P_1>0$ and $P_2>0$, rate splitting with two layers outperforms $R_{\textnormal{M}}(P)$ \eqref{Medard_lower_bound}. In this example, the optimal power allocation is approximately at $P_1 \approx 0.78 P$ and $P_2\approx 0.22 P$. In general, the optimal power allocation is difficult to compute analytically.

\section{Rate-Splitting With $L$ Layers}   \label{sec:L-layering}

One might wonder whether extending our approach to more than two layers can further improve the lower bound. As we shall see in the following section, it does. In fact, for every positive power $P>0$ we show that, once that the power is optimally allocated across layers, the rate-splitting lower bound is strictly increasing in the number of layers.

\subsection{L-Layerings and Rate-Splitting Lower Bounds}
Let $X_1,\ldots,X_L$ be independent, zero-mean, circularly-symmetric, complex Gaussian random variables with respective variances $P_{1},\ldots,P_L$ satisfying
\begin{equation}
\label{power_allocation}
P = \sum_{\ell=1}^L P_{\ell}
\end{equation}
and
\begin{equation}
X_\textnormal{G} = \sum_{\ell=1}^L X_{\ell}.
\end{equation}
Let the cumulative power $Q_{k}$ be given by
\begin{equation}
Q_{k} \triangleq \sum_{\ell=1}^k P_{\ell}.
\end{equation}
We denote the collection of cumulative powers as
\begin{align}
	\BQ \triangleq \bigl\{Q_1,\dotsc,Q_L\bigr\}
\end{align}
and refer to it as an $L$-layering.

It follows from the chain rule for mutual information that
\begin{align}   \label{chain_rule}
	I\bigl(X^L;Y|\hat{H}\bigr)
	= \sum_{\ell=1}^L I\bigl(X_\ell;Y|X^{\ell-1},\hat{H}\bigr)
\end{align}
where we use the shorthand $A^N$ to denote the sequence $A_1,\ldots,A_N$, and $A^0$ denotes the empty sequence. Applying Proposition~\ref{pro:Medard_bound} by replacing the respective $A$, $B$, $C$, and $S$ with
\begin{equation*}
	A \gets \hat{H}, \quad B \gets \hat{H}\sum_{\ell' \neq \ell} X_{\ell'} + \tilde{H}X + Z, \quad C \gets X^{\ell-1}, \quad S \gets X_\ell
\end{equation*}
and by noting that these random variables satisfy the proposition's conditions, we can lower-bound the $\ell$-th summand on the RHS of \eqref{chain_rule} as
\begin{align}    \label{R_l}  
	I\bigl(X_\ell;Y|X^{\ell-1},\hat{H}\bigr)
	&\geq \Exp\left[\log\bigl(1+\Gamma_{\ell,\BQ}(X^{\ell-1},\hat{H})\bigr)\right] \nonumber\\
	&\triangleq R_\ell[\BQ]
\end{align}
where
\begin{align}   \label{Gamma_l}
	\Gamma_{\ell,\BQ}(X^{\ell-1},\hat{H})
	\triangleq \frac{|\hat{H}|^2 P_\ell}{\tilde{V}(\hat{H})\bigl|\sum\limits_{i<\ell} X_i\bigr|^2 + \tilde{V}(\hat{H}) P_\ell + \bigl(|\hat{H}|^2 + \tilde{V}(\hat{H})\bigr) \sum\limits_{i>\ell} P_i + N_0}
\end{align}
and where the last line in \eqref{R_l} should be viewed as the definition of $R_\ell[\BQ]$.
Defining
\begin{equation}
R[\BQ] \triangleq R_1[\BQ]+\ldots+R_{L}[\BQ]
\end{equation}
we obtain from \eqref{chain_rule} and \eqref{R_l} the lower bound
\begin{align}
	I(X_\textnormal{G};Y|\hat{H}) = I\bigl(X^L;Y|\hat{H}\bigr) \geq R[\BQ].
\end{align}

Note that $Q_{\ell-1}=Q_{\ell}$ implies $P_{\ell}=0$, which in turn implies $R_{\ell}[\BQ]=0$. Without loss of optimality, we can therefore restrict ourselves to $L$-layerings satisfying
\begin{equation}   \label{Lfold_strict}
	0 < Q_1< \ldots < Q_L = P.
\end{equation}
We shall denote the set of all $L$-layerings satisfying \eqref{Lfold_strict} by $\mathcal{Q}(P,L)$. Note that this definition of layerings precludes $P=0$, and we shall from now on assume that $P>0$.

Let $R^{\star}(P,L)$ denote the lower bound $R[\BQ]$ optimized over all $\BQ\in\mathcal{Q}(P,L)$, i.e.,
\begin{equation}   \label{R_L}
	R^{\star}(P,L) \triangleq \sup_{\BQ\in\mathcal{Q}(P,L)} R[\BQ].
\end{equation}
In the following, we show that $R^{\star}(P,L)$ is monotonically increasing in $L$. To this end, we need the following lemma.

\begin{mylem}   \label{lem:layering_extension}
Let $L' > L$, and let the $L$-layering $\BQ\in\mathcal{Q}(P,L)$ and the $L'$-layering $\BQ'\in\mathcal{Q}(P,L')$ satisfy
\begin{equation}   \label{extension}
	\bigl\{Q_1,\ldots, Q_L\bigr\} \subset \bigl\{Q'_1,\ldots,Q'_{L'}\bigr\}.
\end{equation}
Then
\begin{equation}
	R[\BQ] \leq R[\BQ']
\end{equation}
with equality if, and only if, $\Prob\bigl\{\hat{H} \cdot \tilde{V}(\hat{H}) = 0\bigr\} = 1$.
\end{mylem}

\begin{IEEEproof}
See Appendix~\ref{app:lem:proof:layering_extension}.
\end{IEEEproof}

\begin{mythm}   \label{cor:layer_monotonicity}
The rate $R^{\star}(P,L)$ is monotonically nondecreasing in $L$. Moreover, if $\Prob\bigl\{\hat{H} \cdot \tilde{V}(\hat{H}) = 0\bigr\} = 1$, then $R^{\star}(P,L)=R_{\textnormal{M}}(P)$ for every $L=1,2\ldots$
\end{mythm}
\begin{IEEEproof}
For every $L$-layering $\BQ\in\mathcal{Q}(P,L)$, we can construct an $(L+1)$-layering $\BQ'\in\mathcal{Q}(P,L+1)$ satisfying $\BQ\subset\BQ'$ by adding $(Q_1+Q_2)/2$ to $\BQ$. Together with Lemma~\ref{lem:layering_extension}, this implies that for every $\BQ\in\mathcal{Q}(P,L)$ there exists a $\BQ'\in\mathcal{Q}(P,L+1)$ such that $R[\BQ]\leq R[\BQ']$, from which we obtain that $R^{\star}(P,L)$ is monotonically nondecreasing upon maximizing both sides of the inequality over all layerings $\BQ\in\mathcal{Q}(P,L)$ and $\BQ'\in\mathcal{Q}(P,L+1)$, respectively.

To show that if $\Prob\bigl\{\hat{H} \cdot \tilde{V}(\hat{H}) = 0\bigr\} = 1$ then $R^{\star}(P,L)=R_{\textnormal{M}}(P)$, $L\in\mathbb{N}$ (where $\mathbb{N}$ denotes the set of positive integers), we first note that M\'edard's lower bound \eqref{Medard_lower_bound} corresponds to $R[\BQ]$ with $\BQ\in\mathcal{Q}(P,1)$. Since the only 1-layering is $\{P\}$, it follows that $R[\BQ]=R^{\star}(P,1)=R_{\textnormal{M}}(P)$. Furthermore, every $L$-layering $\BQ'\in\mathcal{Q}(P,L)$, $L>1$ satisfies $\BQ\subset \BQ'$, so applying Lemma~\ref{lem:layering_extension} with the condition $\Prob\bigl\{\hat{H} \cdot \tilde{V}(\hat{H}) = 0\bigr\} = 1$ yields $R[\BQ']=R[\BQ]=R_{\textnormal{M}}(P)$ for every $\BQ'\in\mathcal{Q}(P,L)$ and $L\in\mathbb{N}$. The claim follows then by maximizing $R[\BQ']$ over all $L$-layerings $\mathcal{Q}(P,L)$.
\end{IEEEproof}

It follows from Theorem~\ref{cor:layer_monotonicity} that the best lower bound, optimized over all layerings of fixed sum-power $P$, namely
\begin{equation}   \label{R_sup}
	R^{\star}(P)
	\triangleq \sup_{L\in\mathbb{N}} \; \sup_{\BQ\in\mathcal{Q}(P,L)} R[\BQ]
	= \sup_{L\in\mathbb{N}} R^{\star}(P,L)
\end{equation}
is approached by letting the number of layers $L$ tend to infinity. An explicit expression for $R^{\star}(P)$ is provided by the following theorem.

\begin{mythm}   \label{thm:infinite_layering}
For a given input power $P$, the supremum of all rate-splitting lower bounds $R[\BQ]$ over \mbox{$\BQ \in \mathcal{Q}(P,L)$} and $L \in \mathbb{N}$ is given by
\begin{align}   \label{R_bar}
	R^{\star}(P)
	&= \lim_{L\to\infty} R^{\star}(P,L) \nonumber\\
	&= \Exp\left[\frac{|\hat{H}|^2}{|\hat{H}|^2 + \tilde{V}(\hat{H}) + \frac{N_0}{P}} \Theta\left( \frac{\tilde{V}(\hat{H})(W-1) - |\hat{H}|^2}{|\hat{H}|^2 + \tilde{V}(\hat{H}) + \frac{N_0}{P}} \right)\right]
\end{align}
where
\begin{align}   \label{Theta}
	\Theta(x) \triangleq \left\{\begin{array}{ll} \frac{1}{x}\log(1+x),\quad & \textnormal{if $-1<x<0$ or $x>0$} \\[3pt] 1, \quad & \textnormal{if $x=0$} \end{array}\right.
\end{align}
and where $W$ is independent of $\hat{H}$ and exponentially distributed with mean $1$.
\end{mythm}
\begin{IEEEproof}
See Appendix~\ref{sec:proof:thm:infinite_layering}.
\end{IEEEproof}

It can be shown that $\Theta(\cdot)$ is a convex function on $(-1,\infty)$, so one can readily recover M\'edard's bound $R_{\textnormal{M}}(P)$ by lower-bounding \eqref{R_bar} via Jensen's inequality.
\begin{myrmk}
The proof of Theorem~\ref{thm:infinite_layering} hinges on the observation that the supremum $R^{\star}(P)$ is approached by an equi-power layering
\begin{align}
	\BU(P,L)
	\triangleq \left\{ \frac{P}{L}, 2\frac{P}{L},\dotsc,(L-1)\frac{P}{L},P \right\}
\end{align}
when the number of layers $L$ is taken to infinity. While this layering was chosen for mathematical convenience, any other layering would also do, provided that some regularity conditions are met. For example, one can show that for any Lipschitz-continuous monotonic bijection $F\colon[0,P]\to [0,P]$, we have
\begin{align}
	\lim_{L \rightarrow \infty} R\bigl[F\bigl(\BU(P,L)\bigr)\bigr]
	= \lim_{L \rightarrow \infty} R\bigl[\BU(P,L)\bigr]
	= R^{\star}(P)
\end{align}
where $F\bigl(\BU(P,L)\bigr)=\bigl\{F(P/L),F(2P/L),\ldots,F(P)\bigr\}$.
\end{myrmk}

\subsection{Upper Bounds}
To assess the tightness of the derived lower bounds, we consider two upper bounds on the Gaussian-input mutual information. The first upper bound is the capacity when the receiver has perfect CSI [cf.~\eqref{coherent_capacity}] and follows by noting that improving the CSI at the receiver does not reduce mutual information:
\begin{equation}
\label{eq:coherent_UB}
I(X_\textnormal{G};Y|\hat{H}) \leq \Exp\left[\log\left(1+\frac{|H|^2 P}{N_0}\right)\right] \triangleq C_\textnormal{coh}(P).
\end{equation}
The second upper bound is given by
\begin{align}   \label{I_upper}
	I(X_\textnormal{G};Y|\hat{H})
	&\leq R_\textnormal{M}(P) + \Exp\left[\log\left(\frac{\tilde{V}(\hat{H})P + N_0}{\tilde{\Phi}(\hat{H})PW+N_0}\right)\right] \nonumber\\
	&\triangleq I_\text{upper}(P)
\end{align}
where $W$ is independent of $\hat{H}$ and is exponentially distributed with mean $1$, and where $\tilde{\Phi}(\hat{h})$ denotes the conditional entropy power of $\tilde{H}$, conditioned on $\hat{H}=\hat{h}$:\footnote{We define $h(\tilde{H}|\hat{H}=\hat{h})=-\infty$ if the conditional distribution of $\tilde{H}$, conditioned on $\hat{H}=\hat{h}$, is not absolutely continuous with respect to the Lebesgue measure.}
\begin{align}
	\tilde{\Phi}(\hat{h})
	\triangleq \left\{\begin{array}{ll} \displaystyle \frac{1}{\pi\e} \e^{h(\tilde{H}|\hat{H}=\hat{h})}, \quad & \textnormal{if $h(\tilde{H}|\hat{H}=\hat{h})>-\infty$}\\[5pt] 0, \quad & \textnormal{otherwise.} \end{array}\right. \label{eq:entropypower}
\end{align}
This upper bound follows from expanding the mutual information as $h(Y|\hat{H}) - h(Y|X_\textnormal{G},\hat{H})$, upper-bounding $h(Y|\hat{H})$ by the entropy of a Gaussian variable of same variance, and lower-bounding $h(Y|X_\textnormal{G},\hat{H})$ using the entropy-power inequality \cite[Theorem~6]{DeCo91}. Using the fact that the Gaussian distribution maximizes differential entropy for a given second moment and that for such a distribution the entropy power equals the variance, it can be shown that
\begin{equation}
\label{eq:EPvsV}
\tilde{\Phi}(\hat{h}) \leq \tilde{V}(\hat{h}), \quad \hat{h}\in\mathbb{C}
\end{equation}
for every conditional distribution of $\tilde{H}$ given $\hat{H}=\hat{h}$ with conditional variance $\tilde{V}(\hat{h})$.

The upper bound \eqref{I_upper} was previously used, e.g., in \cite[Equation~(42)]{BaFoMe01} and\cite[Lemma~2]{YoGo06} for Gaussian fading, in which case \eqref{eq:EPvsV} is tight and the entropy power equals the conditional variance.

\subsection{Numerical Examples}
\begin{figure}[ht]
	\centering
	\subfigure[Bounds vs.\ SNR.]{
		\includegraphics{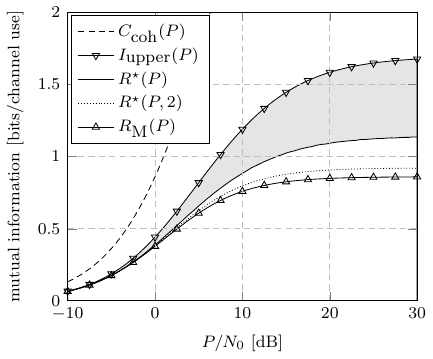}
		\label{fig:R_vs_SNR:fixed_CSI}
	}
	\subfigure[Bounds vs.\ energy per information bit.]{
		\includegraphics{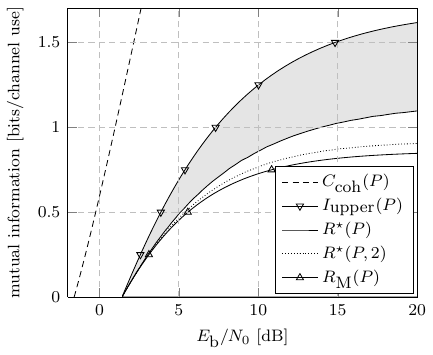}
		\label{fig:energy_per_bit:fixed_CSI}
	}
\caption{Comparison of capacity and Gaussian-input mutual information bounds for fixed CSI.}
\end{figure}

In Figure~\ref{fig:R_vs_SNR:fixed_CSI}, several bounds on the Gaussian-input mutual information $I(X_\textnormal{G};Y|\hat{H})$ are plotted against the SNR on a range from $-10\,\dB$ to $30\,\dB$. From top to bottom, we have the coherent capacity \eqref{eq:coherent_UB}; the upper bound \eqref{I_upper}; the supremum $R^{\star}(P)$ over all rate-splitting bounds (Theorem~\ref{thm:infinite_layering}); the two-layer rate-splitting bound with optimized power allocation $R^{\star}(P,2)$; and M\'edard's lower bound $R_\textnormal{M}(P)$. The grey-shaded area indicates the region in which the curve of the exact Gaussian-input mutual information $I(X_\textnormal{G};Y|\hat{H})$ is located. For this simulation, we have chosen $\hat{H}$ and $\tilde{H}$ to be independent and complex circularly-symmetric Gaussian with parameters $\mu=0$, $\hat{V}=\frac{1}{2}$, and $\tilde{V}(\hat{h}) = \frac{1}{2}$, $\hat{h}\in\mathbb{C}$. Observe that the proposed rate-splitting approach sharpens the bound most at high SNR. In this simulation, the increase $R^{\star}(P)-R_\textnormal{M}(P)$ is approximately $0.28$ bits per channel use as $P$ tends to infinity.

Figure~\ref{fig:energy_per_bit:fixed_CSI} shows the same bounds as Figure~\ref{fig:R_vs_SNR:fixed_CSI}, but this time with the rate plotted against the energy per information bit $E_\textnormal{b}/N_0$. Observe that the minimum energy per bit of all bounds (except that of the coherent capacity $C_\textnormal{coh}$) is equal to $1.41$dB, thus demonstrating that the rate-splitting approach sharpens the bound only marginally at low SNR.

\section{Asymptotically Perfect CSI}   \label{sec:asymptotically_optimal_csi}

The numerical example considered in the previous section (see Figure~\ref{fig:R_vs_SNR:fixed_CSI}) assumes that $\tilde{V}(\hat{H})$ and $\hat{H}$ do not depend on the SNR $P/N_0$. However, in practical communication systems, the channel estimation error---as measured by the mean error variance $\Exp[\tilde{V}(\hat{H})]$---typically decreases as the SNR increases. In this section, we investigate the high-SNR behavior of the derived bounds when $\Exp[\tilde{V}(\hat{H})]$ vanishes as the SNR tends to infinity. When this condition is satisfied, we shall say that we have {\em asymptotically perfect CSI}.

\subsection{Asymptotic Tightness}
We will consider a family of joint distributions of $(\hat{H},\tilde{H})$ parametrized by $\rho=P/N_0$. To make this dependence on $\rho$ explicit, we shall write in this section the two channel components as $\hat{H}_{\rho}$ and $\tilde{H}_{\rho}$, and the respective variances as $\hat{V}_{\rho}$ and $\tilde{V}_{\rho}(\hat{H}_{\rho})$. Similarly, we shall write the entropy power, defined in \eqref{eq:entropypower}, as $\tilde{\Phi}_{\rho}(\hat{H}_{\rho})$. We further adapt the notation to express M\'edard's lower bound, the rate-splitting lower bounds \eqref{R_L} and \eqref{R_sup}, and the upper bounds \eqref{eq:coherent_UB} and \eqref{I_upper} as functions of $\rho$, namely, $R_{\textnormal{M}}(\rho)$, $R^{\star}(\rho,L)$, $R^{\star}(\rho)$, $C_{\text{coh}}(\rho)$, and $I_\text{upper}(\rho)$. 

We assume that $H=\hat{H}_{\rho}+\tilde{H}_{\rho}$ does not depend on $\rho$ and is normalized:
\begin{align}
\label{doby_highSNR_normalized}
	\Exp\bigl[|\hat{H}_{\rho}|^2\bigr] + \Exp\bigl[\tilde{V}_{\rho}(\hat{H}_{\rho})\bigr] = 1.
\end{align}
We further assume that the variance of the estimation error $\tilde{H}_{\rho}$ is not larger than the variance of $H$, i.e., $\tilde{V}_{\rho}(\hat{h}_{\rho}) \leq 1$ for every $\hat{h}_{\rho}\in\mathbb{C}$.

\begin{mythm}   \label{thm:high_SNR}
Let $\hat{H}_{\rho}$, $\tilde{V}_{\rho}(\hat{H}_{\rho})$, and $\tilde{\Phi}_{\rho}(\hat{H}_{\rho})$ satisfy
\begin{subequations}
\begin{IEEEeqnarray}{rCl}
\lim_{\rho \rightarrow \infty} \Exp\bigl[\tilde{V}_{\rho}(\hat{H}_{\rho})\bigr] & = & 0 \label{eq:thm_Vtilde} \\
\varlimsup_{\rho \rightarrow \infty} \left\{ \sup_{\xi\in\mathbb{C}} \frac{\tilde{V}_{\rho}(\xi)}{\tilde{\Phi}_{\rho}(\xi)} \right\} & \leq & M  \label{technical_condition}
\end{IEEEeqnarray}
\end{subequations}
for some finite constant $M$, where we define $0/0\triangleq 1$ and $a/0\triangleq \infty$ for every $a>0$. Then, we have
\begin{align}   \label{statement_of_high_SNR_theorem}
	\varlimsup_{\rho \rightarrow \infty} \bigl\{I(X_{\textnormal{G}};Y|\hat{H}_{\rho}) - R^{\star}(\rho) \bigr\}
	\leq \log(M) \Prob\{|H|>0\}.
\end{align}
\end{mythm}
\begin{IEEEproof}
See Appendix~\ref{sec:thm:proof:high_SNR}.
\end{IEEEproof}
\begin{myrmk}
The proof of Theorem~\ref{thm:high_SNR} reveals that if $\Prob\{|H|>0\}=1$, then one can strengthen \eqref{statement_of_high_SNR_theorem} by replacing $I(X_{\textnormal{G}};Y|\hat{H}_{\rho})$ by its upper bound $I_{\textnormal{upper}}(\rho)$.
\end{myrmk}

If conditioned on (almost) every $\hat{H}_{\rho}=\hat{h}_{\rho}$, the estimation error $\tilde{H}_{\rho}$ is Gaussian, then we have $\tilde{V}_{\rho}(\hat{h}_{\rho})=\tilde{\Phi}_{\rho}(\hat{h}_{\rho})$ for every $\hat{h}_{\rho}\in\mathbb{C}$ and \eqref{technical_condition} is satisfied for $M=1$. Thus, for a conditionally Gaussian $\tilde{H}_{\rho}$, the lower bound $R^{\star}(\rho)$ is asymptotically tight. 
\begin{mycor}
\label{cor:asymptotic_tightness}
Conditioned on every $\hat{H}_{\rho}=\hat{h}_{\rho}$, let $\tilde{H}_{\rho}$ be Gaussian, and let \eqref{eq:thm_Vtilde} and \eqref{technical_condition} hold.
Then, we have
\begin{align}   \label{asymptotic_tightness}
	\lim_{\rho \rightarrow \infty} \bigl\{ I_{\textnormal{upper}}(\rho) - R^{\star}(\rho) \bigr\}
	= 0.
\end{align}
\end{mycor}
\begin{IEEEproof}
The Gaussian distribution of $\tilde{H}_{\rho}$ implies that the cumulative distribution function of $|H|=|\hat{H}_{\rho}+\tilde{H}_{\rho}|$ is continuous, so $\Prob\{H=0\}=0$. The result follows then from \eqref{to_be_bounded} and \eqref{to_be_proven} in the proof of Theorem~\ref{thm:high_SNR} (Appendix~\ref{sec:thm:proof:high_SNR}) upon noting that, for a Gaussian distribution, \eqref{technical_condition} is satisfied for $M=1$.
\end{IEEEproof}
Corollary~\ref{cor:asymptotic_tightness} demonstrates that, for conditionally Gaussian $\tilde{H}_{\rho}$ and asymptotically perfect CSI, both bounds $I_{\text{upper}}(\rho)$ and $R^{\star}(\rho)$ are asymptotically tight in the sense that their difference to the Gaussian-input mutual information vanishes as $\rho$ tends to infinity.  In \cite{LaSh02}, it was argued that the difference between $R_{\textnormal{M}}(\rho)$ and $C_{\textnormal{coh}}(\rho)$ vanishes as $\rho$ tends to infinity if $\tilde{V}_{\rho}(\hat{H}_{\rho})$ decays {\em faster} than the reciprocal of $\rho$, in which case M\'edard's lower bound is asymptotically tight, too. Note however that, if $\tilde{H}_{\rho}$ is conditionally Gaussian, then the upper bound \eqref{I_upper} becomes
\begin{equation}
	I_{\text{upper}}(\rho) = R_{\text{M}}(\rho) + \Exp\left[\log\left(\frac{1+\rho\tilde{V}_{\rho}(\hat{H}_{\rho})}{1+\rho W\tilde{V}_{\rho}(\hat{H}_{\rho})}\right)\right]
\end{equation}
from which follows that
\begin{align}
	\lim_{\rho \rightarrow \infty} \bigl\{ I_{\text{upper}}(\rho) - R_{\textnormal{M}}(\rho) \bigr\}
	= 0 \quad \Longleftrightarrow \quad \lim_{\rho\to\infty} \rho \Exp\bigl[\tilde{V}_{\rho}(\hat{H}_{\rho})] = 0.
\end{align}
Thus, for conditionally Gaussian $\tilde{H}_{\rho}$ and asymptotically perfect CSI, M\'edard's lower bound is asymptotically tight if, and only if, $\Exp\bigl[\tilde{V}_{\rho}(\hat{H}_{\rho})\bigr]$ decays faster than the reciprocal of $\rho$, whereas $R^{\star}(\rho)$ is asymptotically tight {\em irrespective} of the rate of decay.

It follows directly from \eqref{to_be_bounded}--\eqref{Sigma_two_arguments} and Lemma~\ref{lem:g} used within the proof of Theorem~\ref{thm:high_SNR} (Appendix~\ref{sec:thm:proof:high_SNR}) that for any fading distribution satisfying \eqref{technical_condition},
\begin{equation}   \label{gap_upper_bound}
	\varlimsup_{\rho \rightarrow \infty} \bigl\{ I_{\text{upper}}(\rho) - R_{\text{M}}(\rho) \bigr\}
	\leq \gamma + \log(M)
\end{equation}
where $\gamma\approx 0.577$ denotes Euler's constant. Consequently, at high SNR, the bounds $I_\textnormal{upper}(\rho)$, $R^\star(\rho)$, and $R_\textnormal{M}(\rho)$ have all the same logarithmic slope. 

\subsection{Prediction- and Interpolation-Based Channel Estimation}
\label{sec:simulations}

We evaluate the lower bounds $R_{\textnormal{M}}(\rho)$, $R^{\star}(\rho,2)$, and $R^{\star}(\rho)$ together with the upper bound $I_{\text{upper}}(\rho)$ for two specific channel estimation errors satisfying \eqref{eq:thm_Vtilde}. We assume that $\hat{H}_{\rho}$ and $\tilde{H}_{\rho}$ are zero-mean, circularly-symmetric, complex Gaussian random variables that are independent of each other\footnote{Consequently, $\tilde{V}_{\rho}$ does not depend on $\hat{H}_{\rho}$ either.} and satisfy the normalization \eqref{doby_highSNR_normalized}. The former has variance $\hat{V}_{\rho}$ and the latter has variance $\tilde{V}_{\rho}$. We consider variances $\tilde{V}_{\rho}$ of the forms
\begin{subequations}
\begin{IEEEeqnarray}{lCl}
\tilde{V}_{\rho} & = & \left( \frac{1}{2B} + \frac{1}{\rho} \right)^{2B} \rho^{2B-1} - \frac{1}{\rho} \label{prediction_error_2} 
\end{IEEEeqnarray}
and
\begin{IEEEeqnarray}{lCl}
\tilde{V}_{\rho} & = &\frac{2 B T}{\rho + 2 B T} \label{interpolation_error}
\end{IEEEeqnarray}
\end{subequations}
for some $0<B<\frac{1}{2}$, where $T=\lfloor1/(2B)\rfloor$ is the largest integer not greater than $1/(2B)$.

As we shall argue next, \eqref{prediction_error_2} corresponds to prediction-based channel estimation, whereas \eqref{interpolation_error} corresponds to interpolation-based channel estimation. Indeed, suppose for a moment that the fading process $\{H[k]\}_{k\in\mathbb{Z}}$ is not i.i.d.\ (as assumed in Section~\ref{introduction}) but is a zero-mean, unit-variance, stationary, circularly-symmetric, complex Gaussian process with power spectral density
\begin{align}   \label{rectangular_density}
	f_{H}(\lambda)
	=
	\begin{cases}
		\displaystyle\frac{1}{2B}, & |\lambda| < B \\
		0,            & B \leq |\lambda| \leq \frac{1}{2}
	\end{cases}
\end{align}
for some $0 < B < \frac{1}{2}$. The fading's autocovariance function is determined by $f_H(\cdot)$ through the expression
\begin{equation}
\Exp\bigl[H[k+m](H[k])^*\bigr] = \int_{-1/2}^{1/2} \e^{\mathsf{i} 2\pi m \lambda} f_{H}(\lambda) \intd \lambda
\end{equation}
where $(\cdot)^*$ denotes complex conjugation and $\mathsf{i}\triangleq\sqrt{-1}$. 

We obtain \eqref{prediction_error_2} if we let $\hat{H}[k]$ be the minimum mean-square error (MMSE) predictor in predicting $H[k]$ from a noisy observation of its past
\begin{equation}
H[k-1]\sqrt{P} + Z[k-1],H[k-2]\sqrt{P} + Z[k-2],\ldots
\end{equation}
Indeed, in this case $\hat{H}[k]$ and $\tilde{H}[k]=H[k]-\hat{H}[k]$ are zero-mean, circularly-symmetric, complex Gaussian random variables that are independent of each other, the latter with mean zero and variance \cite[Section~10.8,~p.~181--184]{GrSz84},\cite[Equation~(11)]{La05}
\begin{align}   \label{prediction_error}
	\tilde{V}_{\rho}
	= \exp\left\{ \int_{-1/2}^{1/2} \log\left( f_{H}(\lambda) + \frac{1}{\rho} \right) \intd \lambda \right\} - \frac{1}{\rho}.
\end{align}
For the power spectral density \eqref{rectangular_density} this gives \eqref{prediction_error_2}. Note that, even though the lower bounds $R_{\text{M}}(\rho)$, $R^{\star}(\rho,L)$, and $R^{\star}(\rho)$ were derived for i.i.d.\ fading $\{\hat{H}_{\rho}[k],\tilde{H}_{\rho}[k]\}_{k\in\mathbb{Z}}$, by evaluating them for $\tilde{H}_{\rho}[k]$ having variance \eqref{prediction_error_2}, they can be used to derive lower bounds on the capacity of noncoherent fading channels with stationary fading having power spectral density $f_H(\cdot)$; see, e.g., \cite{La05}.

The variance \eqref{interpolation_error} corresponds to a channel-estimation scheme where the transmitter emits every $T$ time instants (say at $k=nT$, $n\in\mathbb{Z}$) a pilot symbol $\sqrt{P}$ and where the receiver estimates the fading coefficients at the remaining time instants $k$ (i.e., where $k$ is not an integer multiple of $T$) from the noisy observations
\begin{equation}
H[nT]\sqrt{P} + Z[nT], \quad n\in\mathbb{Z}
\end{equation}
using an MMSE interpolator; see, e.g., \cite{DoToSa04,Lo08,AsKoGu11,AsKoGu13}. When the power spectral density $f_H(\cdot)$ is bandlimited to $B$ and when $T\leq 1/(2B)$, it can be shown that the variance of the estimation error is given by \cite{OhGi02}
\begin{align} 
	\tilde{V}_{\rho}
	&= 1 - \int_{-B}^B \frac{\rho f_{H}^2(\lambda)}{\rho f_{H}(\lambda) + T} \intd \lambda.
\end{align}
For the power spectral density \eqref{rectangular_density} this gives \eqref{interpolation_error}. Again, even though the lower bounds $R_{\text{M}}(\rho)$, $R^{\star}(\rho,L)$, and $R^{\star}(\rho)$ were derived for i.i.d.\ fading $\{\hat{H}_{\rho}[k],\tilde{H}_{\rho}[k]\}_{k\in\mathbb{Z}}$, by evaluating them for $\tilde{H}_{\rho}[k]$ having variance \eqref{interpolation_error}, they can be directly used to derive lower bounds on the capacity of noncoherent fading channels with stationary fading having power spectral density $f_H(\cdot)$, provided that we account for the rate loss due to the transmission of pilots. In fact, it was shown that, when $1/(2B)$ is an integer, the above interpolation-based channel estimation scheme together with M\'edard's lower bound $R_{\text{M}}(\rho)$ achieves the capacity pre-log \cite{Lo08,AsKoGu11,AsKoGu13}.\footnote{The \emph{capacity pre-log} is defined as the limiting ratio of the capacity to $\log(\rho)$ as $\rho$ tends to infinity. In multiple-input multiple-output (MIMO) systems, it is sometimes also referred to as the \emph{number of degrees of freedom} or the \emph{multiplexing gain}.}

\subsection{Numerical Examples}

For Figures~\ref{fig:prediction}--\ref{fig:R_vs_SNR:interpolation:relative} below, we assume that $\hat{H}_{\rho}$ and $\tilde{H}_{\rho}$ are independent, zero-mean, circularly-symmetric, complex Gaussian random variables.

Figure~\ref{fig:R_vs_SNR:prediction} shows the lower bounds $R_{\text{M}}(\rho)$, $R^{\star}(\rho,2)$, and $R^{\star}(\rho)$ together with the upper bounds $I_{\text{upper}}(\rho)$ and $C_{\text{coh}}(\rho)$ as a function of $\rho$ for $\tilde{H}_{\rho}$ having variance \eqref{prediction_error_2}, with $B=1/4$. Figure~\ref{fig:energy_per_bit:prediction} shows the same bounds, but as a function of the energy per information bit. The curve of the exact Gaussian-input mutual information $I(X_\textnormal{G};Y|\hat{H})$ is located within the shaded area. Observe that, in contrast to the curves in Figure~\ref{fig:R_vs_SNR:fixed_CSI}, all curves are unbounded in the SNR, which is a consequence of the fact that $\tilde{V}_{\rho}$ vanishes as $\rho$ tends to infinity. Further observe that the shaded area narrows down as $\rho$ grows. This is consistent with Corollary~\ref{cor:asymptotic_tightness}, which states that for (conditionally) Gaussian $\tilde{H}_{\rho}$ and asymptotically perfect CSI, the bounds $I_{\textnormal{upper}}(\rho)$ and $R^{\star}(\rho)$ are asymptotically tight. Note that, as demonstrated by \eqref{gap_upper_bound}, the upper bound $I_{\text{upper}}(\rho)$ and all lower bounds have the same logarithmic slope at high SNR.

\begin{figure}
	\centering
	\subfigure[Bounds vs.\ SNR.]{
		\includegraphics{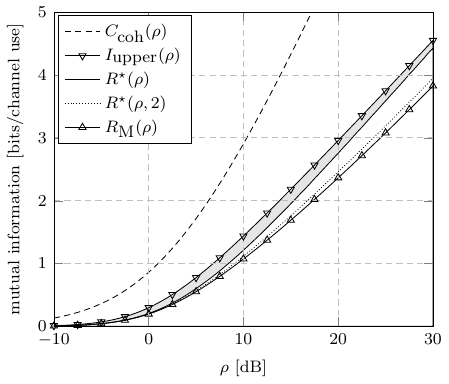}
		\label{fig:R_vs_SNR:prediction}
	}
	\subfigure[Bounds vs.\ energy per information bit.]{
		\includegraphics{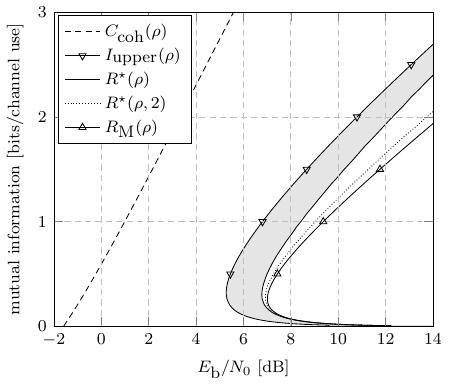}
		\label{fig:energy_per_bit:prediction}
	}
	\caption{Prediction-based channel estimation.}
	\label{fig:prediction}
\end{figure}

\begin{figure}[htb]
	\centering
	\subfigure[Bounds vs.\ SNR.]{
		\includegraphics{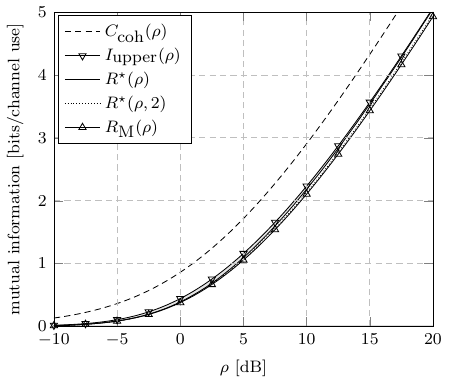}
		\label{fig:R_vs_SNR:interpolation}
	}
	\subfigure[Bounds vs.\ energy per information bit.]{
		\includegraphics{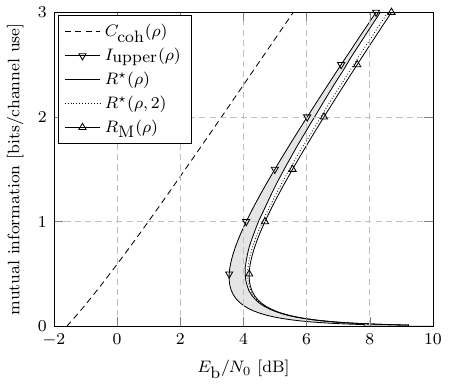}
		\label{fig:energy_per_bit:interpolation}
	}
	\caption{Interpolation-based channel estimation.}
	\label{fig:interpolation}
\end{figure}
\begin{figure}[h]
\begin{center}
	\includegraphics{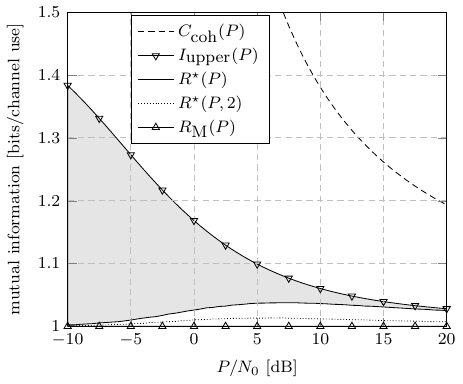}
	\caption{Bounds from Figure~\ref{fig:R_vs_SNR:interpolation}, divided by $R_\textnormal{M}(P)$.}
	\label{fig:R_vs_SNR:interpolation:relative}
\end{center}
\end{figure}

Figure~\ref{fig:R_vs_SNR:interpolation} shows the lower bounds $R_{\text{M}}(\rho)$, $R^{\star}(\rho,2)$, and $R^{\star}(\rho)$ together with the upper bounds $I_{\text{upper}}(\rho)$ and $C_{\text{coh}}(\rho)$ as a function of $\rho$ for $\tilde{H}_{\rho}$ having variance \eqref{interpolation_error}, with $BT=1/2$. Again, observe that all curves are unbounded in the SNR and that the lower bound $R^{\star}(\rho)$ is asymptotically tight as $\rho$ tends to infinity. What is more, $R^{\star}(\rho)$ is close to $I_{\text{upper}}(\rho)$ for a large range of SNR. Further observe that, at high SNR, the upper bound $I_{\text{upper}}(\rho)$ and all lower bounds have the same logarithmic slope as $C_{\text{coh}}(\rho)$. This fact was used in \cite{Lo08,AsKoGu11,AsKoGu13} to derive tight lower bounds on the capacity pre-log of noncoherent fading channels.

Figure~\ref{fig:R_vs_SNR:interpolation:relative} shows the same plots as Figure~\ref{fig:R_vs_SNR:interpolation}, except that all values have been divided by $R_\textnormal{M}(\rho)$ so as to visualize the {\em relative} improvement of the rate-splitting bounds over M\'edard's bound. We observe that, at low SNR, these improvements are negligible. This indicates that the rate-splitting bounds may be more interesting at moderate and high SNR than at low SNR.

\section{Relationship to Mismatched Decoding}\label{sec:mismatch}

We have demonstrated that M\'edard's lower bound $R_{\text{M}}(P)$ on the capacity of fading channels with imperfect CSI can be sharpened by using a rate-splitting approach: by expressing the Gaussian input $X_{\textnormal{G}}$ as the sum of $L$ Gaussian random variables $X_{1},\ldots,X_L$, by applying the chain rule for mutual information to express $I(X;Y|\hat{H})$ as
\begin{equation}
\label{eq:last_chain_rule}
I(X;Y|\hat{H}) = \sum_{\ell=1}^L I\bigl(X_{\ell};Y\bigm|X^{\ell-1},\hat{H}\bigr)
\end{equation}
and by lower-bounding each mutual information on the RHS of \eqref{eq:last_chain_rule} using M\'edard's bounding technique, we obtain a lower bound that is strictly larger than $R_\text{M}(P)$.

This result is reminiscent of a result in the mismatched decoding literature. Indeed, it has been shown that, if $\tilde{V}(\hat{h})$ is independent of $\hat{h}$, then M\'edard's lower bound $R_{\text{M}}(P)$ is the \emph{generalized mutual information (GMI)}\footnote{For a given channel and decoding rule, the GMI is the rate below which the average probability of error---averaged over the ensemble of i.i.d.\ codebooks---decays to zero as the blocklength tends to infinity, and above which this average tends to one.} \cite{KaSh93,Sti66,GaLaTe00} of the above channel \eqref{system_equation} when the codebook is drawn according to an i.i.d.\ Gaussian distribution and when the decoding rule is the \emph{scaled nearest neighbor decoding rule} under which the decoder chooses the message $m$ that minimizes \cite[Corollary~3.0.1]{LaSh02}
\begin{equation}
D(m) = \frac{1}{n} \sum_{k=1}^n \Bigl|y[k]-\hat{h}[k] \, x^{(m)}[k]\Bigr|^2.
\end{equation}
Here, $\left(x^{(m)}[1],\ldots,x^{(m)}[n]\right)$ denotes the codeword associated with the message $m\in\{1,\ldots,\lceil e^{n R}\rceil\}$ (where $\lceil x\rceil$ denotes the smallest integer not smaller than $x$), and $R$ and $n$ denote the rate and the blocklength of the code, respectively. It has been further shown that, for a given decoding rule, treating the single-user channel as a multiple-access channel (MAC) can sometimes yield an achievable rate that is larger than the GMI or other achievable rates corresponding to codebooks under which the codewords are drawn independently \cite{La96}.\footnote{It is unknown whether this is true for the scaled nearest-neighbor decoding rule.}

Since the above rate-splitting approach treats the single-user channel \eqref{system_equation} as an $L$-user MAC with channel inputs $X_{1},\ldots,X_{L}$, i.e.,
\begin{equation}
\label{eq:MAC}
Y = \sum_{\ell=1}^L (\hat{H}+\tilde{H})X_{\ell} + Z
\end{equation}
it may therefore seem plausible that this approach can sharpen M\'edard's lower bound. Note, however, that it is unknown whether $R[\BQ]$ can be achieved with a scaled nearest-neighbor decoder or a modified version thereof.\footnote{In opposition to what we wrote in \cite{PaKoFo12}, the modified scaled nearest-neighbor decoder presented in \cite[Equations~(23) and (24)]{PaKoFo12} achieves a GMI not larger than $R_{\text{M}}(P)$.}

\section{Summary and Conclusion}\label{sec:summary}

We have demonstrated that rate splitting can increase the well-known capacity lower bound \eqref{Medard_lower_bound} by M\'edard \cite{Me00} of fading channels with imperfect channel-state information at the receiver. By computing the supremum of these bounds over all possible rate-splitting strategies, we have established a novel capacity lower bound which is larger than M\'edard's lower bound \eqref{Medard_lower_bound}.

We have further studied the high-SNR behavior of the novel bound under the assumption that the variance of the channel estimation error tends to zero with the SNR. We have shown that, for a Gaussian estimation error, the rate-splitting bound is asymptotically tight in the sense that its difference to the Gaussian-input mutual information vanishes as the SNR tends to infinity. In contrast to M\'edard's lower bound, which is asymptotically tight only if the variance of the estimation error decays faster than the reciprocal of the SNR, the novel lower bound is asymptotically tight irrespective of the rate at which this variance decays.

While the novel rate-splitting bound outperforms M\'edard's bound, one may argue that it is less practical due to the successive decoding strategy, which is more susceptible to error propagation. Nevertheless, we believe that the rate-splitting bound has theoretical importance, since it may be useful in finding the capacity of noncoherent fading channels. For example, as mentioned in the previous paragraph for asymptotically perfect CSI, the rate-splitting bound converges to the Gaussian-input mutual information as the SNR tends to infinity. Consequently, at high SNR, any gap to capacity is merely due to the (potentially suboptimal) Gaussian input distribution and not due to the bounding techniques used to evaluate mutual information. In order to find the high-SNR capacity of this channel, it thus remains to assess the optimality of Gaussian inputs. While such inputs are highly suboptimal for imperfect CSI, they may in fact be optimal when the CSI is asymptotically perfect.

\appendices

\section{Proof of Proposition~\ref{pro:Medard_bound}}   \label{app:pro:proof:Medard_bound}

We expand the mutual information as
\begin{align}
	I(S;AS+B|A,C)
	= h(S|A,C) - h(S|AS+B,A,C).\label{eq:app_1}
\end{align}
Since, by assumption, $S$ is zero-mean, variance-$P$, circularly-symmetric, complex Gaussian and independent of $(A,C)$, the first entropy on the RHS of \eqref{eq:app_1} is readily evaluated as
\begin{equation}
h(S|A,C) = h(S) = \log(\pi\e P). \label{eq:app1_Gauss}
\end{equation}
Conditioned on $(A,C)=(a,c)$, the second entropy can be upper-bounded as follows:
\begin{IEEEeqnarray}{lCl}
h(S|AS+B,A=a,C=c) & = & h\bigl(S-\alpha_{A,C}(AS+B-\mu_{B|A,C})\bigm|AS+B,A=a,C=c\bigr) \nonumber\\
& \leq & h\bigl(S-\alpha_{A,C}(AS+B-\mu_{B|A,C})\bigm|A=a,C=c\bigr) \nonumber\\
& \leq & \log\left(\pi e \Exp\Bigl[\bigl|S-\alpha_{A,C}(AS+B-\mu_{B|A,C})\bigr|^2\Bigm| A=a,C=c\Bigr]\right) \label{eq:app1_cond}
\end{IEEEeqnarray}
for any arbitrary $\alpha_{a,c}\in\mathbb{C}$, where $\mu_{B|a,c}\triangleq\Exp[B|A=a,C=c]$. Here the first inequality follows because conditioning cannot increase entropy, and the second inequality follows from the entropy-maximizing property of the Gaussian distribution. Combining \eqref{eq:app1_cond} with \eqref{eq:app1_Gauss} and \eqref{eq:app_1} thus yields for every $(A,C)=(a,c)$ and $\alpha_{a,c}$
\begin{equation}
I(S;AS+B|A=a,C=c) \geq \log\frac{P}{\Exp\left[\left.|S-\alpha_{A,C}(AS+B-\mu_{B|A,C})|^2\right| A=a,C=c\right]}. \label{eq:app_2}
\end{equation}

We choose $\alpha_{a,c}$ so that $\alpha_{a,c}(aS+B-\mu_{B|a,c})$ is the linear MMSE estimate of $S$ from the observation $AS+B$ given $(A,C)=(a,c)$, namely,
\begin{equation}
\label{eq:app_alpha_ac}
\alpha_{a,c} = \frac{\Exp\bigl[S(AS+B-\mu_{B|A,C})^*\bigm|A=a,C=c\bigr]}{\Exp\bigl[|AS+B-\mu_{B|A,C}|^2\bigm|A=a,C=c\bigr]} = \frac{a^*P}{|a|^2 P+V_B(a,c)}
\end{equation}
where $V_B(a,c)$ denotes the conditional variance of $B$ conditioned on $(A,C)=(a,c)$. Here we have used that, conditioned on $(A,C)=(a,c)$, the random variables $S$ and $B$ are uncorrelated and that $S$ has zero mean and variance $P$ and is independent of $(A,C)$. Combining these conditions with \eqref{eq:app_alpha_ac}, we obtain
\begin{equation}
\Exp\Bigl[\bigl|S-\alpha_{A,C}(AS+B-\mu_{B|A,C})\bigr|^2\Bigm| A=a,C=c\Bigr] = P \frac{V_B(a,c)}{|a|^2 P+V_B(a,c)}. \label{eq:app1_variance}
\end{equation}
Consequently, \eqref{eq:app1_variance} and \eqref{eq:app_2} give for every $(A,C)=(a,c)$
\begin{equation}
I(S;AS+B|A=a,C=c) \geq \log\left(1+\frac{|a|^2 P}{V_B(a,c)}\right).
\end{equation}
Proposition~\ref{pro:Medard_bound} follows then by averaging over $(A,C)$.

\section{Proof of Lemma~\ref{lem:layering_extension}}   \label{app:lem:proof:layering_extension}
To prove Lemma~\ref{lem:layering_extension}, we shall demonstrate for every $L\in\mathbb{N}$ that, if the layerings $\BQ\in\mathcal{Q}(P,L)$ and $\BQ'\in\mathcal{Q}(P,L+1)$ satisfy
\begin{equation}
	\bigl\{Q_1,\ldots,Q_L\bigr\} \subset \bigl\{Q_1',\ldots,Q'_{L+1}\bigr\}
\end{equation}
then $R[\BQ]\leq R[\BQ']$ with equality if, and only if, $\Prob\bigl\{\hat{H} \cdot \tilde{V}(\hat{H}) = 0\bigr\} = 1$. The general case where $\BQ'\in\mathcal{Q}(P,L')$ for some arbitrary $L'>L$ follows directly from the case $L'=L+1$ by applying the above result $(L'-L)$ times.

Let the element in $\BQ'$ that is not contained in $\BQ$ be at position $\tau\in\{1,\ldots,L\}$, i.e.,\footnote{By the definition of a layering, we have $Q'_{L+1}=Q_L=P$, so the element in $\BQ'$ not contained in $\BQ$ cannot be at position $\tau=L+1$.}
\begin{subequations}
\begin{equation}
\label{iota1}
Q_{\ell} = Q'_{\ell}, \quad \ell=1,\ldots,\tau-1
\end{equation}
and
\begin{equation}
\label{iota2}
Q_{\ell} = Q'_{\ell+1}, \quad \ell=\tau,\ldots,L.
\end{equation}
\end{subequations}
We next express $\Gamma_{\ell,\BA}(X^{\ell-1},\hat{H})$ in \eqref{Gamma_l} for some general layering $\BA$ as
\begin{align}
	\Gamma_{\ell,\BA}(X^{\ell-1},\hat{H})
	= \frac{|\hat{H}|^2 (A_{\ell}-A_{\ell-1})}{\tilde{V}(\hat{H})\left|\sum_{i<\ell}X_i\right|^2 + \bigl(|\hat{H}|^2+\tilde{V}(\hat{H})\bigr)P - |\hat{H}|^2 A_{\ell}-\tilde{V}(\hat{H})A_{\ell-1} + N_0}.
\end{align}
Noting that for the layering $\BQ$ the term $\left|\sum_{i<\ell}X_i\right|^2$ has an exponential distribution with mean $Q_{\ell-1}$, whereas for the layering $\BQ'$ it has an exponential distribution with mean $Q_{\ell-1}'$, and using \eqref{iota1} and \eqref{iota2}, it can be easily verified that
\begin{align}
	\Exp\left[\log\bigl(1+\Gamma_{\ell,\BQ}(X^{\ell-1},\hat{H})\bigr)\right]
	= \Exp\left[\log\bigl(1+\Gamma_{\ell,\BQ'}(X^{\ell-1},\hat{H})\bigr)\right], \quad \ell=1,\ldots,\tau-1
\end{align}
and
\begin{align}
	\Exp\left[\log\bigl(1+\Gamma_{\ell,\BQ}(X^{\ell-1},\hat{H})\bigr)\right]
	= \Exp\left[\log\bigl(1+\Gamma_{\ell+1,\BQ'}(X^{\ell},\hat{H})\bigr)\right], \quad \ell=\tau+1,\ldots,L.
\end{align}
Subtracting $R[\BQ]$ from $R[\BQ']$ yields
\begin{IEEEeqnarray}{lCl}
	R[\BQ'] - R[\BQ]
	& = & \Exp\left[\log\bigl(1+\Gamma_{\tau,\BQ'}(X^{\tau-1},\hat{H})\bigr)\right] \nonumber\\
	& & {} + \Exp\left[\log\bigl(1+\Gamma_{\tau+1,\BQ'}(X^{\tau},\hat{H})\bigr)\right]
	- \Exp\left[\log\bigl(1+\Gamma_{\tau,\BQ}(X^{\tau-1},\hat{H})\bigr)\right]. \label{RQ-RQ'}
\end{IEEEeqnarray}
Since the random variables $X_1,\ldots,X_{\tau},\hat{H}$ are independent, we can express the second expectation as
\begin{equation}   \label{two_exps}
	\Exp\left[\left.\Exp_{X_{\tau}}\left[\log\bigl(1+\Gamma_{\tau+1,\BQ'}(X^{\tau},\hat{H})\bigr)\right|X^{\tau-1},\hat{H}\right]\right]
\end{equation}
where the subscript indicates that the inner expected value is computed with respect to $X_{\tau}$. Using that, for every $\alpha>0$, the function $x \mapsto \log(1+\alpha/x)$ is strictly convex in $x>0$, it follows from Jensen's inequality that, for every $X^{\tau-1}=x^{\tau-1}$ and $\hat{H}=\hat{h}$, the inner expectation is lower-bounded by\footnote{With a slight abuse of notation, we write $\Gamma_{\tau+1,\BQ'}(x^{\tau},\hat{h})$ as $\Gamma_{\tau+1,\BQ'}(x^{\tau-1},x_{\tau},\hat{h})$.}
\begin{align}   \label{Upsbar_Jensen}
	\Exp_{X_{\tau}}\left[\log\bigl(1+\Gamma_{\tau+1,\BQ'}(x^{\tau-1},X_{\tau},\hat{h})\bigr)\right]
	\geq \log\bigl(1+\bar{\Gamma}_{\tau+1,\BQ'}(x^{\tau-1},\hat{h})\bigr)
\end{align}
where we define
\begin{align}   \label{Gamma_bar}
	\bar{\Gamma}_{\tau+1,\BQ'}(x^{\tau-1},\hat{h})
	\triangleq \frac{|\hat{h}|^2 (Q'_{\tau+1}-Q'_{\tau})}{\tilde{V}(\hat{h})\left|\sum_{i<\tau} x_i\right|^2+(|\hat{h}|^2+\tilde{V}(\hat{h})) P -|\hat{h}|^2 Q'_{\tau+1}-\tilde{V}(\hat{h})Q'_{\tau-1}+N_0}.
\end{align}
The denominator of ${\bar{\Gamma}}_{\tau+1,\BQ'}(x^{\tau-1},\hat{h})$ is obtained by noting that $X_{\tau}$ has zero mean, so
\begin{align}
	\Exp_{X_{\tau}}\left[\left|\sum_{i<\tau} x_i + X_{\tau}\right|^2\right]=\left|\sum_{i<\tau} x_i\right|^2 + Q'_{\tau}-Q'_{\tau-1}.
\end{align}
Since $\BQ'\in\mathcal{Q}(P,L+1)$ implies that $\Exp\left[|X_{\tau}|^2\right]>0$, the inequality in \eqref{Upsbar_Jensen} is strict except in the trivial cases $\tilde{V}(\hat{h})=0$ or $\hat{h}=0$. Combining \eqref{two_exps} and \eqref{Upsbar_Jensen} yields
\begin{align}   \label{Exp_log_Upsbar}   
	\Exp\left[\log\bigl(1+\Gamma_{\tau+1,\BQ'}(X^{\tau},\hat{H})\bigr)\right]
	&\geq \Exp\left[\log\bigl(1+\bar{\Gamma}_{\tau+1,\BQ'}(X^{\tau-1},\hat{H})\bigr)\right]
\end{align}
which together with \eqref{RQ-RQ'} gives
\begin{align}   \label{LB_Upsbar}   
	R[\BQ']-R[\BQ]
	&\geq \Exp\left[\log\bigl(1+\Gamma_{\tau,\BQ'}(X^{\tau-1},\hat{H})\bigr)\right]
		+ \Exp\left[\log\left(\frac{1+\bar{\Gamma}_{\tau+1,\BQ'}(X^{\tau-1},\hat{H})}{1+\Gamma_{\tau,\BQ}(X^{\tau-1},\hat{H})}\right)\right]
\end{align}
with the inequality being strict except if $\Prob\bigl\{\hat{H} \cdot \tilde{V}(\hat{H})=0\bigr\}=1$.

We next use \eqref{iota1} and \eqref{iota2} and the fact that $\left|\sum_{i<\tau} X_i\right|^2$ has an exponential distribution with mean $Q'_{\tau-1}$ under both layerings $\BQ$ and $\BQ'$ to evaluate the second expected value on the RHS of \eqref{LB_Upsbar}:
\begin{align}
	\Exp\left[\log\left(\frac{1+\bar{\Gamma}_{\tau+1,\BQ'}(X^{\tau-1},\hat{H})}{1+\Gamma_{\tau,\BQ}(X^{\tau-1},\hat{H})}\right)\right]
	&= \Exp\left[\log\left(\frac{T-|\hat{H}|^2 Q'_{\tau}-\tilde{V}(\hat{H}) Q'_{\tau-1}+N_0}{T - |\hat{H}|^2 Q'_{\tau-1}-\tilde{V}(\hat{H})Q'_{\tau-1}+N_0}\right)\right] \nonumber\\
	&= -\Exp\left[\log\left(1+\frac{|\hat{H}|^2(Q'_{\tau}-Q'_{\tau-1})}{T -|\hat{H}|^2Q'_{\tau}-\tilde{V}(\hat{H})Q'_{\tau-1}+N_0}\right)\right]
\end{align}
where we introduce
\begin{equation}
T \triangleq \tilde{V}(\hat{H})\left|\sum_{i<\tau} X_i\right|^2 +\bigl(|\hat{H}|^2+\tilde{V}(\hat{H})\bigr)P
\end{equation}
for ease of exposition. By noting that
\begin{equation}
\frac{|\hat{H}|^2(Q'_{\tau}-Q'_{\tau-1})}{T -|\hat{H}|^2Q'_{\tau}-\tilde{V}(\hat{H})Q'_{\tau-1}+N_0} = \Gamma_{\tau,\BQ'}(X^{\tau-1},\hat{H})
\end{equation}
it follows that the RHS of \eqref{LB_Upsbar} is zero, thus demonstrating that
\begin{equation}
R[\BQ] \leq R[\BQ']
\end{equation}
with equality if, and only if, $\Prob\bigl\{\hat{H} \cdot \tilde{V}(\hat{H})=0\bigr\}=1$. This proves Lemma~\ref{lem:layering_extension}.

\section{}
\label{app:III}
\subsection{Proof of Theorem~\ref{thm:infinite_layering}}   \label{sec:proof:thm:infinite_layering}
To prove Theorem~\ref{thm:infinite_layering}, we first note that for $\hat{h}=0$
\begin{equation}
\log\left(1+\frac{|\hat{h}|^2P}{\tilde{V}(\hat{h})P+N_0}\right) = 0
\end{equation}
whereas for $\tilde{V}(\hat{h})=0$
\begin{equation}
\log\left(1+\frac{|\hat{h}|^2P}{\tilde{V}(\hat{h})P+N_0}\right) = \log\left(1+\frac{|\hat{h}|^2P}{N_0}\right)
\end{equation}
which in both cases is equal to
\begin{equation}
\frac{|\hat{h}|}{|\hat{h}|+\tilde{V}(\hat{h})+\frac{N_0}{P}} \Theta\left( \frac{\tilde{V}(\hat{h})(W-1) - |\hat{h}|^2}{|\hat{h}|^2 + \tilde{V}(\hat{h}) + \frac{N_0}{P}} \right)
\end{equation}
where $W$ is as in Theorem~\ref{thm:infinite_layering}. This implies that, if $\Prob\bigl\{\hat{H} \cdot \tilde{V}(\hat{H}) = 0\bigr\} = 1$, then
\begin{equation}
R_{\text{M}}(P) = \Exp\left[\frac{|\hat{H}|}{|\hat{H}|+\tilde{V}(\hat{H})+\frac{N_0}{P}} \Theta\left( \frac{\tilde{V}(\hat{H})(W-1) - |\hat{H}|^2}{|\hat{H}|^2 + \tilde{V}(\hat{H}) + \frac{N_0}{P}} \right)\right]
\end{equation}
from which Theorem~\ref{thm:infinite_layering} follows because, by Theorem~\ref{cor:layer_monotonicity}, $R^{\star}(P,L)=R_{\textnormal{M}}(P)$, $L\in\mathbb{N}$.

In the following, we consider the case where $\Prob\bigl\{\hat{H} \cdot \tilde{V}(\hat{H}) = 0\bigr\} < 1$. To this end, we first show that it suffices to consider equi-power layerings
\begin{align}   \label{uniform_layering}
	\BU(P,K)
	\triangleq \left\{ \frac{P}{K}, 2\frac{P}{K},\dotsc,(K-1)\frac{P}{K},P \right\}.
\end{align}
More precisely, we shall show that for every $L$-layering $\BQ\in\mathcal{Q}(P,L)$ there exists some sufficiently large $K$ such that $\BU(P,K)$ outperforms $\BQ$, i.e.,
\begin{align}   \label{uniform_eventually_wins}
	R[\BU(P,K)] > R[\BQ].
\end{align}
This then implies that
\begin{equation}
	R^{\star}(P)
	= \sup_{L\in\mathbb{N}} \left\{ \sup_{\BQ\in\mathcal{Q}(P,L)} R[\BQ] \right\}
	= \sup_{K\in\mathbb{N}} R[\BU(P,K)]
\end{equation}
from which we obtain, by Lemma~\ref{lem:layering_extension}, that
\begin{equation}   \label{uniform_lim}
	R^{\star}(P) = \varlimsup_{K\to\infty} R[\BU(P,K)]
\end{equation}
upon noting that $\BU(P,K)\subset\BU(P,2K)$ for every $K\in\mathbb{N}$.

To show that for every $L$-layering $\BQ\in\mathcal{Q}(P,L)$ there exists some $\BU(P,K)$ (with $K$ sufficiently large) outperforming $\BQ$, we first note that for every $\epsilon>0$ one can find a sufficiently large $K$ and two $(L+1)$-layerings $\BS\in\mathcal{Q}(P,L+1)$ and $\BT\in\mathcal{Q}(P,L+1)$ satisfying $\BQ\subset\BS$ and $\BT\subset\BU(P,K)$ such that
\begin{equation}
\label{vanishing_approximation_error}
\max_{1\leq \ell \leq L+1} |T_{\ell} - S_{\ell}| \leq \epsilon.
\end{equation}
Indeed, $\BS$ may be obtained by including $(Q_1+Q_2)/2$ into $\BQ$, i.e., $\BS = \BQ \cup \{(Q_1+Q_2)/2\}$. Furthermore, for $K$ larger than $P/(\min_{0\leq\ell\leq L}|S_{\ell+1}-S_{\ell}|)$ (where $S_0=0$ by convention), choosing
\begin{equation*}
T_{\ell} = \left\lceil\frac{S_{\ell} K}{P}\right\rceil \frac{P}{K}, \quad \ell=1,\ldots,L+1
\end{equation*}
yields $\BT\subset\BU(P,K)$ and
\begin{equation}
\max_{1\leq \ell \leq L+1} |T_{\ell} - S_{\ell}| \leq \frac{P}{K}
\end{equation}
from which \eqref{vanishing_approximation_error} follows. To prove \eqref{uniform_eventually_wins}, we then need the following lemma.

\begin{mylem}   \label{lem:continuity}
The function $R[\BQ]$ satisfies
\begin{equation}
	\lim_{\BQ \to \BQ'} R[\BQ] = R[\BQ']
\end{equation}
where $\BQ\to\BQ'$ is to be understood as $\max_{\ell} |Q_{\ell}-Q_{\ell}'| \to 0$ with $\BQ$ and $\BQ'$ having an equal number of layers.
\end{mylem}

\begin{IEEEproof}
See Appendix~\ref{app:lem:proof:continuity}.
\end{IEEEproof}

From Lemma~\ref{lem:continuity} and from the observation \eqref{vanishing_approximation_error}, it follows that for every $\delta > 0$ there exists a sufficiently large $K$ such that 
\begin{equation}
\bigl|R[\BT]-R[\BS]\bigr| \leq \delta.
\end{equation}
Since by Lemma~\ref{lem:layering_extension} and the assumption $\Prob\bigl\{\hat{H} \cdot \tilde{V}(\hat{H}) = 0\bigr\} < 1$ we have
\begin{equation}
R[\BQ] < R[\BS] \quad \textnormal{and} \quad R[\BT] < R[\BU(P,K)]
\end{equation}
this yields
\begin{equation}
R[\BQ] < R[\BS] \leq R[\BT]+\delta
\end{equation}
which for a sufficiently small $\delta$ is strictly smaller than $R[\BU(P,K)]$ due to $\BT\subset\BU(P,K)$. This proves \eqref{uniform_eventually_wins}.

Recalling that \eqref{uniform_eventually_wins} implies \eqref{uniform_lim}, we continue by evaluating $R[\BU(P,K)]$ in the limit as $K$ tends to infinity. To this end, we write $R[\BU(P,K)]$ as
\begin{align}   \label{sum_rate}
	R[\BU(P,K)] = \sum_{\ell=1}^K \Exp\left[\log\bigl(1 + \Gamma_{\ell,\BU}(W_{\ell},\hat{H})\bigr)\right]
\end{align}
with [cf.~\eqref{Gamma_l}]
\begin{align}
	\Gamma_{\ell,\BU}(W_{\ell},\hat{H})
	&= \frac{|\hat{H}|^2}{\tilde{V}(\hat{H})(\ell-1)W_{\ell} + \tilde{V}(\hat{H}) + \bigl(|\hat{H}|^2 + \tilde{V}(\hat{H})\bigr)(K - \ell) + N_0\frac{K}{P}} \label{Gamma_l_2}
\end{align}
and
\begin{equation}
	W_{\ell} \triangleq \left\{\begin{array}{ll} 0, \quad & \ell=1\\[5pt] \displaystyle \frac{1}{(\ell-1)\frac{P}{K}}\left|\textstyle\sum_{i<\ell} X_i\right|^2, \quad &\ell=2,\ldots,K. \end{array}\right.
\end{equation}
The random variables $(W_1,\ldots,W_K)$ are dependent but have equal marginals. (Each marginal has a unit-mean exponential distribution.) Since the RHS of \eqref{sum_rate} depends on $(W_1,\ldots,W_K)$ only via their marginal distributions, we can thus express $R[\BU(P,K)]$ as
\begin{equation}
\label{uniform_W}
R[\BU(P,K)] = \Exp\left[\sum_{\ell=1}^K \log\bigl(1+\Gamma_{\ell,\BU}(W,\hat{H})\bigr)\right]
\end{equation}
where $W$ is independent of $\hat{H}$ and has a unit-mean exponential distribution.

Combining \eqref{uniform_W} with \eqref{uniform_lim} yields
\begin{align}
	R^{\star}(P)
	&= \varlimsup_{K \to \infty} \Exp\left[ \sum_{\ell=1}^K \log\bigl(1+\Gamma_{\ell,\BU}(W,\hat{H})\bigr)\right].
\end{align}
We next show that
\begin{IEEEeqnarray}{lCl}
	R^{\star}(P) & = & \Exp\left[\lim_{K\to\infty} \sum_{\ell=1}^K \Gamma_{\ell,\BU}(W,\hat{H})\right]
\end{IEEEeqnarray}
and evaluate $\sum_{\ell=1}^K \Gamma_{\ell,\BU}(W,\hat{H})$ for every $(W,\hat{H})=(w,\hat{h})$ in the limit as $K$ tends to infinity. To this end, we first lower-bound $R^{\star}(P)$ using Fatou's Lemma \cite[(1.6.8),~p.~50]{AsDo00} and the lower bound $\log(1+x)\geq x - x^2/2$, $x\geq 0$:
\begin{align}
	R^{\star}(P)
		&= \varlimsup_{K \to \infty} \Exp\left[ \sum_{\ell=1}^K \log\bigl(1+\Gamma_{\ell,\BU}(W,\hat{H})\bigr)\right] \nonumber\\
		&\geq \Exp \left[\varliminf_{K \to \infty} \sum_{\ell=1}^K \log\bigl(1+\Gamma_{\ell,\BU}(W,\hat{H})\bigr) \right] \nonumber\\
		&\geq \Exp\left[ \varliminf_{K \to \infty} \sum_{\ell=1}^K \Gamma_{\ell,\BU}(W,\hat{H})\right] - \frac{1}{2} \Exp \left[ \varlimsup_{K \to \infty} \sum_{\ell=1}^K \Gamma^2_{\ell,\BU}(W,\hat{H}) \right] \label{Fatou}
\end{align}
where $\varliminf$ denotes the \emph{limit inferior}. We next argue that the second term on the RHS of \eqref{Fatou} is zero. Indeed, we have for every $(W,\hat{H})=(w,\hat{h})$ [cf.~\eqref{Gamma_l_2}]
\begin{align}
	\Gamma^2_{\ell,\BU}(w,\hat{h})
	&= \frac{|\hat{h}|^4}{\left[\tilde{V}(\hat{h})(\ell-1)w + \tilde{V}(\hat{h}) + \bigl(|\hat{h}|^2 + \tilde{V}(\hat{h})\bigr)(K - \ell) + N_0\frac{K}{P}\right]^2} \nonumber\\
	&\leq \frac{|\hat{h}|^4}{\left[\min\left\{ \tilde{V}(\hat{h})w , \bigl(|\hat{h}|^2 + \tilde{V}(\hat{h})\bigr) \right\} (K - 1) + \tilde{V}(\hat{h}) + N_0\frac{K}{P} \right]^2}
\end{align}
where the inequality follows from observing that the denominator of $\Gamma^2_{\ell,\BU}(w,\hat{h})$ is the square of a positive affine linear function of $\ell \in \{1,\dots,K\}$ and is therefore minimized for either $\ell=1$ or $\ell=K$. This yields
\begin{align}
	\sum_{\ell=1}^K \Gamma^2_{\ell,\BU}(w,\hat{h})
	&\leq \frac{K |\hat{h}|^4}{\left[\min\left\{ \tilde{V}(\hat{h})w, \bigl(|\hat{h}|^2 + \tilde{V}(\hat{h})\bigr) \right\} (K - 1) + \tilde{V}(\hat{h}) + N_0\frac{K}{P} \right]^2}. \label{upperbound_Fatou}
\end{align}
Since $\sum_{\ell=1}^K \Gamma^2_{\ell,\BU}(w,\hat{h})$ is nonnegative, and since the RHS of \eqref{upperbound_Fatou} vanishes as $K$ tends to infinity, it follows that, for every $(W,\hat{H})=(w,\hat{h})$,
\begin{equation}
\label{Fatou_zero}
\varlimsup_{K\to\infty} \sum_{\ell=1}^K \Gamma^2_{\ell,\BU}(w,\hat{h}) = 0.
\end{equation}
Combining \eqref{Fatou_zero} with \eqref{Fatou} yields
\begin{equation}
\label{Fatou_LB}
R^{\star}(P) \geq \Exp\left[ \varliminf_{K \to \infty} \sum_{\ell=1}^K \Gamma_{\ell,\BU}(W,\hat{H})\right].
\end{equation}
We next show that
\begin{equation}
\label{Fatou_UB}
R^{\star}(P) \leq \Exp\left[ \varlimsup_{K \to \infty} \sum_{\ell=1}^K \Gamma_{\ell,\BU}(W,\hat{H}) \right].
\end{equation}
To this end, we first use the upper bound $\log(1+x)\leq x$, $x\geq 0$ to obtain
\begin{IEEEeqnarray}{lCl}
R^{\star}(P) & = &  \varlimsup_{K \to \infty} \Exp\left[ \sum_{\ell=1}^K \log\bigl(1+\Gamma_{\ell,\BU}(W,\hat{H})\bigr)\right] \nonumber\\
& \leq & \varlimsup_{K \to \infty} \Exp\left[ \sum_{\ell=1}^K \Gamma_{\ell,\BU}(W,\hat{H})\right].
\end{IEEEeqnarray}
Noting that, for every $(W,\hat{H})=(w,\hat{h})$, the sum inside the expectation is upper-bounded by
\begin{equation}
\sum_{\ell=1}^K \Gamma_{\ell,\BU}(w,\hat{h}) \leq \sum_{\ell=1}^K \frac{|\hat{h}|^2}{N_0 \frac{K}{P}} = \frac{P|\hat{h}|^2}{N_0} \triangleq \zeta(\hat{h})
\end{equation}
and noting that, since $\hat{H}$ has a finite second moment, we have that $0<\Exp[\zeta(\hat{H})]<\infty$, we obtain \eqref{Fatou_UB} upon applying Fatou's Lemma to the nonnegative function $(w,\hat{h})\mapsto\zeta(\hat{h})-\sum_{\ell=1}^K \Gamma_{\ell,\BU}(w,\hat{h})$.

It remains to show that, for every $(W,\hat{H})=(w,\hat{h})$,
\begin{align}
	\lim_{K\to\infty} \sum_{\ell=1}^K \Gamma_{\ell,\BU}(w,\hat{h})
	= \frac{|\hat{h}|^2}{|\hat{h}|^2 + \tilde{V}(\hat{h}) + \frac{N_0}{P}} \Theta\left( \frac{\tilde{V}(\hat{h})(w-1) - |\hat{h}|^2}{|\hat{h}|^2 + \tilde{V}(\hat{h}) + \frac{N_0}{P}} \right) \label{limit_expr}
\end{align}
where $\Theta(\cdot)$ is defined in \eqref{Theta}. This then implies that the bounds \eqref{Fatou_LB} and \eqref{Fatou_UB} coincide and
\begin{IEEEeqnarray}{lCl}
R^{\star}(P)
& = & \Exp\left[\frac{|\hat{H}|^2}{|\hat{H}|^2 + \tilde{V}(\hat{H}) + \frac{N_0}{P}} \Theta\left( \frac{\tilde{V}(\hat{H})(W-1) - |\hat{H}|^2}{|\hat{H}|^2 + \tilde{V}(\hat{H}) + \frac{N_0}{P}} \right)\right] \label{Rbarlim}
\end{IEEEeqnarray}
which proves Theorem~\ref{thm:infinite_layering}.

To show \eqref{limit_expr}, we express the denominator in $\Gamma_{\ell,\BU}(w,\hat{h})$ as $a\ell+bK+c$ with
\begin{subequations}
\begin{align}
	a &= \tilde{V}(\hat{h})(w - 1) - |\hat{h}|^2 \\
	b &= |\hat{h}|^2 + \tilde{V}(\hat{h}) + \frac{N_0}{P} \\
	c &= \tilde{V}(\hat{h})(1 - w)
\end{align}
\end{subequations}
allowing us to write
\begin{equation}
\label{Sk_vs_Sktilde}
\sum_{\ell=1}^K \Gamma_{\ell,\BU}(w,\hat{h}) = \sum_{\ell=1}^K\frac{|\hat{h}|^2}{a\ell+bK+c}.
\end{equation}
Observe that, for every $(w,\hat{h})$, $a+b$ and $a$ are strictly positive.

If $a=0$, then we get the limit
\begin{equation}
\label{a=0}
\lim_{K\to\infty} \sum_{\ell=1}^K \Gamma_{\ell,\BU}(w,\hat{h}) = \frac{|\hat{h}|^2}{b}.
\end{equation}
We next consider the case $a\neq 0$. Note that
\begin{equation}
\label{newsum}
\lim_{K\to\infty}\left(\sum_{\ell=1}^K\frac{|\hat{h}|^2}{a\ell+bK+c} - \sum_{\ell=1}^K\frac{|\hat{h}|^2}{a\ell+bK}\right) = 0.
\end{equation}
Indeed, by the triangle inequality, we have
\begin{align}
	\left|\sum_{\ell=1}^K\frac{|\hat{h}|^2}{a\ell+bK+c} - \sum_{\ell=1}^K\frac{|\hat{h}|^2}{a\ell+bK}\right|
	\leq \sum_{\ell=1}^K \frac{|c| |\hat{h}|^2}{(a\ell+bK+c)(a\ell+bK)}.
\end{align}
Since the two factors $(a\ell+bK+c)$ and $(a\ell+bK)$ appearing in the denominator are both positive affine functions of $\ell$ with equal coefficient $a$, their product takes its extremal values at $\ell=1$ or $\ell=K$, depending on the sign of $a$. If $a>0$, then
\begin{align}
\label{eq:bla1}
	\left|\sum_{\ell=1}^K\frac{|\hat{h}|^2}{a\ell+bK+c} - \sum_{\ell=1}^K\frac{|\hat{h}|^2}{a\ell+bK}\right|
	\leq \frac{K|c| |\hat{h}|^2}{(a+bK+c)(a+bK)}.
\end{align}
If $a \leq 0$, then
\begin{align}
\label{eq:bla2}
	\left|\sum_{\ell=1}^K\frac{|\hat{h}|^2}{a\ell+bK+c} - \sum_{\ell=1}^K\frac{|\hat{h}|^2}{a\ell+bK}\right|
	\leq \frac{K|c| |\hat{h}|^2}{((a+b)K+c)(a+b)K}.
\end{align}
Since the RHS of \eqref{eq:bla1} and of \eqref{eq:bla2} vanish as $K$ tends to infinity, this yields \eqref{newsum}. Consequently,
\begin{IEEEeqnarray}{lCl}
\lim_{K\to\infty}\sum_{\ell=1}^K\frac{|\hat{h}|^2}{a\ell+bK+c} & = & \lim_{K\to\infty} \sum_{\ell=1}^K\frac{|\hat{h}|^2}{a\ell+bK} \nonumber\\
& = & \lim_{K\to\infty} \frac{1}{K} \sum_{\ell=1}^K \frac{|\hat{h}|^2}{a\frac{\ell}{K}+b} \nonumber\\
& = & \int_0^1 \frac{|\hat{h}|^2}{a x + b} \intd x \nonumber\\
& = & \frac{|\hat{h}|^2}{a}\log\left(1+\frac{a}{b}\right) \IEEEeqnarraynumspace\label{aneq0}
\end{IEEEeqnarray}
where the third step follows by noting that the function $x \mapsto \frac{1}{ax+b}$ is Riemann integrable, so the Riemann sum converges to the integral. 

Using the definition of $\Theta(\cdot)$ [cf.~\eqref{Theta}], it follows from \eqref{a=0} and \eqref{aneq0} that
\begin{equation}
\lim_{K\to\infty} \sum_{\ell=1}^K \Gamma_{\ell,\BU}(w,\hat{h}) = \frac{|\hat{h}|^2}{b}\Theta\left(\frac{a}{b}\right) = \frac{|\hat{h}|^2}{|\hat{h}|^2 + \tilde{V}(\hat{h}) + \frac{N_0}{P}} \Theta\left( \frac{\tilde{V}(\hat{h})(w-1) - |\hat{h}|^2}{|\hat{h}|^2 + \tilde{V}(\hat{h}) + \frac{N_0}{P}} \right)
\end{equation}
thus proving \eqref{limit_expr}, which in turn proves Theorem~\ref{thm:infinite_layering}.

\subsection{Proof of Lemma~\ref{lem:continuity}}   \label{app:lem:proof:continuity}

We show that
\begin{equation}
\label{app:continuity_lim}
\lim_{\BQ \to \BQ'} R[\BQ] = R[\BQ']
\end{equation}
where $\BQ\to\BQ'$ should be read as
\begin{equation}
\max_{\ell} \bigl|Q_{\ell}-Q_{\ell}'\bigr| \to 0.
\end{equation}
To this end, we write $R[\BQ]$ as
\begin{equation}
\label{app:continuity_R}
R[\BQ] = \sum_{\ell=1}^L \Exp\left[\log\bigl(1+\Gamma_{\ell,\BQ}(W_{\ell},\hat{H})\bigr)\right]
\end{equation}
with
\begin{align}
	\Gamma_{\ell,\BQ}(W_{\ell},\hat{H})
	& \triangleq \frac{|\hat{H}|^2 (Q_{\ell}-Q_{\ell-1})}{\tilde{V}(\hat{H})W_{\ell} Q_{\ell-1} + \tilde{V}(\hat{H})(Q_{\ell}-Q_{\ell-1}) + \bigl(|\hat{H}|^2+\tilde{V}(\hat{H})\bigr)(P-Q_{\ell}) + N_0}
\end{align}
(assuming that $Q_0=0$) and
\begin{equation}
W_{\ell} \triangleq \left\{\begin{array}{ll} 0, \quad & \ell=0\\\displaystyle \frac{1}{Q_{\ell-1}} \left|\sum_{i<\ell} X_i\right|^2, \quad & \ell=2,\ldots,L. \end{array}\right.
\end{equation}
Using that, with probability one,
\begin{equation}
0 \leq \log\bigl(1+\Gamma_{\ell,\BQ}(W_{\ell},\hat{H})\bigr)\leq \frac{|\hat{H}|^2 P}{N_0}
\end{equation}
and that $\hat{H}$ has finite variance, it follows from the Dominated Convergence Theorem \cite[(1.6.9),~p.~50]{AsDo00} that
\begin{align}
	\lim_{\BQ\to\BQ'}  \Exp\left[\log\bigl(1+\Gamma_{\ell,\BQ}(W_{\ell},\hat{H})\bigr)\right]
	&= \Exp\left[\lim_{\BQ\to\BQ'}\log\bigl(1+\Gamma_{\ell,\BQ}(W_{\ell},\hat{H})\bigr)\right] \nonumber\\
	&= \Exp\left[\log\bigl(1+\Gamma_{\ell,\BQ'}(W_{\ell},\hat{H})\bigr)\right] \label{app:continuity_ell}
\end{align}
where the last step follows by noting that, for every $(w_{\ell},\hat{h})$, the function $\BQ\mapsto\log\bigl(1+\Gamma_{\ell,\BQ}(w_{\ell},\hat{h})\bigr)$ is continuous. Combining \eqref{app:continuity_ell} with \eqref{app:continuity_R} proves \eqref{app:continuity_lim} and, hence, Lemma~\ref{lem:continuity}.

\section{}
\label{app:IV}
\subsection{Proof of Theorem~\ref{thm:high_SNR}}   \label{sec:thm:proof:high_SNR}

To prove Theorem~\ref{thm:high_SNR}, we show that, in the limit as the SNR tends to infinity, the difference
\begin{equation}
\label{eq:difference}
I(X_{\textnormal{G}};Y|\hat{H}_{\rho}) - R^{\star}(\rho)
\end{equation}
is upper-bounded by $\log(M)\Prob\{|H|>0\}$ provided that \eqref{eq:thm_Vtilde}--\eqref{technical_condition} are satisfied. To this end, we introduce the random variable
\begin{equation}
D \triangleq \begin{cases} 0 & \text{if $H=0$} \\1 & \text{if $|H|>0$} \end{cases}
\end{equation}
and upper-bound the mutual information in \eqref{eq:difference} as
\begin{equation}
\label{eq:doby_UB_D}
I(X_{\textnormal{G}};Y|\hat{H}_{\rho}) \leq I(X_{\textnormal{G}};Y|\hat{H}_{\rho},D)
\end{equation}
which follows because $X_{\textnormal{G}}$ is independent of $\hat{H}_{\rho}$ and $D$. We next note that
\begin{equation}
\label{eq:doby_D=0}
I(X_{\textnormal{G}};Y|\hat{H}_{\rho},D=0)= I(X_{\textnormal{G}};Z) = 0
\end{equation}
since $X_{\textnormal{G}}$, $Z$, and $(\hat{H}_{\rho},H)$ are independent. If $\Prob\{H=0\}=1$, then Theorem~\ref{thm:high_SNR} follows directly from \eqref{eq:doby_UB_D}, \eqref{eq:doby_D=0}, and the nonnegativity of $R^{\star}(\rho)$. In the following, we assume that $\Prob\{H=0\}<1$.

We express $R^{\star}(\rho)$ in \eqref{R_bar} as $\Exp[R^{\star}(\rho,W,\hat{H}_{\rho})]$ with\footnote{Recall $W$ is independent of $\hat{H}_{\rho}$ and has a unit-mean exponential distribution.}
\begin{equation}
R^{\star}(\rho,w,\xi) \triangleq \frac{|\xi|^2}{|\xi|^2 + \tilde{V}_{\rho}(\xi) + \rho^{-1}} \Theta\left( \frac{(w-1)\tilde{V}_{\rho}(\xi) - |\xi|^2}{|\xi|^2 + \tilde{V}_{\rho}(\xi) + \rho^{-1}} \right), \quad \bigl(\rho>0,w\geq 0, \xi\in\mathbb{C}\,\bigr). \label{R_bar_three_arguments}
\end{equation}
Note that, by the definition of $\Theta(\cdot)$ in \eqref{Theta}, $R^{\star}(\rho,w,\xi)\geq 0$ for every $(\rho>0,w\geq 0, \xi\in\mathbb{C})$. Using this result together with \eqref{eq:doby_UB_D} and \eqref{eq:doby_D=0}, we obtain
\begin{equation}
I(X_{\textnormal{G}};Y|\hat{H}_{\rho}) - R^{\star}(\rho) \leq I(X_{\textnormal{G}};Y|\hat{H}_{\rho},D=1) \Prob\{|H|>0\} - \Exp\bigl[R^{\star}(\rho,W,\hat{H}_{\rho})\bigm| D=1\bigr] \Prob\{|H|>0\}.
\end{equation}
To prove Theorem~\ref{thm:high_SNR}, it remains to show that if \eqref{eq:thm_Vtilde}--\eqref{technical_condition} hold, then
\begin{equation}
\label{eq:doby_2bound_D=1}
\varlimsup_{\rho\to\infty} \left\{ I(X_{\textnormal{G}};Y|\hat{H}_{\rho},D=1) - \Exp\bigl[R^{\star}(\rho,W,\hat{H}_{\rho})\bigm| D=1\bigr]\right\} \leq \log(M).
\end{equation}
For ease of exposition, we will omit in the remainder of the proof the conditioning on the event $|H|>0$ and replace tacitly the joint distribution of $(\hat{H}_{\rho},H)$ by its conditional distribution, conditioned on $|H|>0$. This change will not affect the bounds \eqref{Medard_lower_bound}, \eqref{R_bar}, and \eqref{I_upper}, since they hold irrespective of the distribution of $(\hat{H}_{\rho},H)$ (provided that $H$ and $\hat{H}_{\rho}$ satisfy the conditions indicated in Section~\ref{introduction}). Note that, under this new distribution, we have $\Prob\{H=0\}=0$.

To prove \eqref{eq:doby_2bound_D=1}, we upper-bound $I(X_{\textnormal{G}};Y|\hat{H}_{\rho})$ by $I_{\text{upper}}(\rho)$ using \eqref{I_upper} and express $I_{\text{upper}}(\rho) - R^{\star}(\rho)$ as
\begin{IEEEeqnarray}{lCl}
I_{\text{upper}}(\rho) - R^{\star}(\rho) & = & \Exp\bigl[R_\textnormal{M}(\rho,\hat{H}_{\rho})] + \Exp\bigl[\Delta(\rho,W,\hat{H}_{\rho})\bigr] - \Exp\bigl[R^{\star}(\rho,W,\hat{H}_{\rho})\bigr] \nonumber\\
& = & \Exp\bigl[ \Sigma(\rho,\hat{H}_{\rho})\bigr] \label{to_be_bounded}
\end{IEEEeqnarray}
where we have defined [cf.~\eqref{Medard_lower_bound}, \eqref{I_upper}]
\begin{IEEEeqnarray}{rCll}
R_{\textnormal{M}}(\rho,\xi)
	&\triangleq & \log\left(1 + \frac{|\xi|^2}{\tilde{V}_{\rho}(\xi) + \rho^{-1}}\right), \quad & \bigl(\rho>0,w\geq 0, \xi\in\mathbb{C}\,\bigr)   \label{R_M_two_arguments} \\
\Delta(\rho,w,\xi)
	&\triangleq & \log\left(\frac{\tilde{V}_{\rho}(\xi) + \rho^{-1}}{\tilde{\Phi}_{\rho}(\xi)w + \rho^{-1}}\right), \quad & \bigl(\rho>0,w\geq 0, \xi\in\mathbb{C}\,\bigr)   \label{Delta_three_arguments}
\end{IEEEeqnarray}
and
\begin{align} 
	\Sigma(\rho,\xi)
	&\triangleq R_{\textnormal{M}}(\rho,\xi) + \Exp\bigl[\Delta(\rho,W,\xi)\bigr] - \Exp\bigl[R^{\star}(\rho,W,\xi)\bigr], \quad \bigl(\rho>0,\,\xi\in\mathbb{C}\bigr).   \label{Sigma_two_arguments}
\end{align}
Note that $\Sigma(\rho,\xi)\geq0$,  $\xi\in\mathbb{C}$ since $I_{\text{upper}}(\rho) - R^{\star}(\rho)$ is nonnegative for any distribution of $\hat{H}_{\rho}$, hence it is also nonnegative if $\hat{H}_{\rho}=\xi$ with probability one.

We next show that
\begin{align}   \label{to_be_proven}
	\varlimsup_{\rho \rightarrow \infty} \Exp\bigl[ \Sigma(\rho,\hat{H}_{\rho})\bigr]
	\leq \log(M).
\end{align}
To this end, we write the RHS of \eqref{to_be_bounded} as
\begin{IEEEeqnarray}{lCl}
	\Exp\bigl[ \Sigma\bigl(\rho,\hat{H}_{\rho}\bigr)\bigr]
	& = & \Exp\Bigl[ \Sigma\bigl(\rho,\hat{H}_{\rho}\bigr) \operatorname{I}\bigl\{|\hat{H}_{\rho}| \leq \xi_0\bigr\} \Bigr] + \Exp\Bigl[ \Sigma\bigl(\rho,\hat{H}_{\rho}\bigr) \operatorname{I}\bigl\{|\hat{H}_{\rho}| > \xi_0\bigr\} \Bigr]    \label{Sigma_limit_2}
\end{IEEEeqnarray}
for some arbitrary $0<\xi_0<1$, where $\operatorname{I}\{\cdot\}$ denotes the indicator function (it is $1$ if the statement in the curly brackets is true and is $0$ otherwise). We then show that
\begin{subequations}
\begin{equation}
\lim_{\xi_0\downarrow 0}\varlimsup_{\rho\to\infty}\Exp\Bigl[ \Sigma\bigl(\rho,\hat{H}_{\rho}\bigr) \operatorname{I}\bigl\{|\hat{H}_{\rho}| \leq \xi_0\bigr\} \Bigr] = 0 \label{eq:doby_Sigma1}
\end{equation}
and
\begin{equation}
\varlimsup_{\xi_0\downarrow 0}\varlimsup_{\rho\to\infty}\Exp\Bigl[ \Sigma\bigl(\rho,\hat{H}_{\rho}\bigr) \operatorname{I}\bigl\{|\hat{H}_{\rho}| > \xi_0\bigr\} \Bigr] \leq \log(M). \label{eq:doby_Sigma2}
\end{equation}
\end{subequations}
To prove \eqref{eq:doby_Sigma1}, we need the following two lemmas.
\begin{mylem}   \label{lem:g}
We have
\begin{align}
	\varlimsup_{\rho\to\infty}\sup_{\xi\in\mathbb{C}}\Sigma(\rho,\xi)
	\leq \gamma + \log(M).
\end{align}
where $\gamma\approx 0.577$ denotes Euler's constant.
\end{mylem}
\begin{IEEEproof}
See Appendix~\ref{app:lem:proof:g}.
\end{IEEEproof}
\begin{mylem}   \label{lem:estimation_convergence}
	Let $\tilde{V}_{\rho}(\hat{H}_{\rho})$ and $\tilde{H}_{\rho}$ satisfy \eqref{eq:thm_Vtilde} and \eqref{technical_condition}, and assume that $\Prob\{H=0\}=0$. Then
\begin{align}
	\lim_{\xi_0\downarrow 0}\varliminf_{\rho \rightarrow \infty} \Prob\bigl\{ |\hat{H}_{\rho}| > \xi_0\bigr\} = 1.
\end{align}
\end{mylem}
\begin{IEEEproof}
See Appendix~\ref{app:lem:proof:estimation_convergence}.
\end{IEEEproof}
Lemma~\ref{lem:g} implies that for every $\epsilon>0$ there exists a $\rho_0>0$ such that
\begin{equation}
\sup_{\xi\in\mathbb{C}}\Sigma(\rho,\xi) \leq \gamma + \log(M)+\epsilon, \quad \rho \geq \rho_0.
\end{equation}
Consequently, for $\rho \geq \rho_0$ we have
\begin{IEEEeqnarray}{lCl}
\Exp\Bigl[ \Sigma\bigl(\rho,\hat{H}_{\rho}\bigr) \operatorname{I}\bigl\{|\hat{H}_{\rho}| \leq \xi_0\bigr\}\Bigr]
& \leq & \bigl(\gamma + \log(M)+\epsilon\bigr)\Prob\bigl\{ |\hat{H}_{\rho}| \leq \xi_0\bigr\}. \IEEEeqnarraynumspace \label{eq:doby_Sigma1_beforelim}
\end{IEEEeqnarray}
Together with Lemma~\ref{lem:estimation_convergence}, this yields \eqref{eq:doby_Sigma1} upon taking limits on both sides of \eqref{eq:doby_Sigma1_beforelim}:
\begin{IEEEeqnarray}{lCl}
\IEEEeqnarraymulticol{3}{l}{\lim_{\xi_0\downarrow 0}\varlimsup_{\rho\to\infty} \left\{ \Exp\Bigl[ \Sigma\bigl(\rho,\hat{H}_{\rho}\bigr) \operatorname{I}\bigl\{|\hat{H}_{\rho}| \leq \xi_0\bigr\}\Bigr] \right\}} \nonumber\\
\quad & \leq & \bigl(\gamma+\log(M)+\epsilon\bigr) \left\{\lim_{\xi_0\downarrow 0}\varlimsup_{\rho\to\infty} \Prob\bigl\{ |\hat{H}_{\rho}| \leq \xi_0\bigr\}\right\} \nonumber\\
& = & 0.
\end{IEEEeqnarray}

To prove \eqref{eq:doby_Sigma2}, we first upper-bound $\Sigma(\rho,\xi)$ by lower-bounding $\Exp[R^{\star}(\rho,W,\xi)]$ for $\rho>0$ and $|\xi|>\xi_0$ using that $R^{\star}(\rho,w,\xi)$ is nonnegative and recalling that $W$ is unit-mean exponentially distributed:
\begin{align}
	\Exp\bigl[R^{\star}(\rho,W,\xi)\bigr]
	\geq \int_0^{\kappa(\rho,\xi)} R^{\star}(\rho,w,\xi) \e^{-w} \intd w, \quad \bigl(\rho>0,\,|\xi|>\xi_0\bigr) \label{eq:doby_Sigma2_LB1}
\end{align}
where
\begin{align}   \label{kappa}
	\kappa(\rho,\xi)
	\triangleq \frac{\xi_0^2}{\sqrt{\tilde{V}_{\rho}(\xi)+\rho^{-1}}}.
\end{align}
This choice for $\kappa(\rho,\xi)$ together with the assumption $\tilde{V}_{\rho}(\xi)\leq 1$ ensures that $(1-w)\tilde{V}_{\rho}(\xi) + |\xi|^2$ is strictly positive for all values of the integration variable $w$ and for all $|\xi|>\xi_0$. Using \eqref{R_bar_three_arguments} and the definition \eqref{Theta} of the function $\Theta(\cdot)$, the lower bound \eqref{eq:doby_Sigma2_LB1} reads as
\begin{align}   \label{R_bar_lower_bound_2}
	\Exp\bigl[R^{\star}(\rho,W,\xi)\bigr]
	&\geq \int_0^{\kappa(\rho,\xi)} \frac{|\xi|^2}{(w-1)\tilde{V}_{\rho}(\xi) - |\xi|^2} \log\left( 1 + \frac{(w-1)\tilde{V}_{\rho}(\xi) - |\xi|^2}{|\xi|^2 + \tilde{V}_{\rho}(\xi) + \rho^{-1}} \right) \e^{-w} \intd w, \quad \bigl(\rho>0,\,|\xi|>\xi_0\bigr).
\end{align}
Combining \eqref{R_bar_lower_bound_2} with \eqref{R_M_two_arguments}--\eqref{Sigma_two_arguments} yields
\begin{align}   \label{Sigma_upper_bound_1}
	\Sigma(\rho,\xi)
	&\leq \log\left(1 + \frac{|\xi|^2}{\tilde{V}_{\rho}(\xi) + \rho^{-1}}\right) + \int_0^\infty \log\left(\frac{\tilde{V}_{\rho}(\xi) + \rho^{-1}}{\tilde{\Phi}_{\rho}(\xi)w + \rho^{-1}}\right) \e^{-w} \intd w \nonumber\\
	&\qquad {} + \int_0^{\kappa(\rho,\xi)} \frac{|\xi|^2}{(1-w)\tilde{V}_{\rho}(\xi) + |\xi|^2} \log\left( 1 + \frac{(w-1)\tilde{V}_{\rho}(\xi) - |\xi|^2}{|\xi|^2 + \tilde{V}_{\rho}(\xi) + \rho^{-1}} \right) \e^{-w} \intd w, \quad \bigl(\rho>0,\,|\xi|>\xi_0\bigr).
\end{align}
This upper bound has the form $\Sigma(\rho,\xi) \leq \Sigma_1(\rho,\xi) + \Sigma_2(\rho,\xi) + \Sigma_3(\rho,\xi)$ where the terms can be expanded as
\begin{subequations}   \label{Sigma_i}
\begin{IEEEeqnarray}{rCl}
	\Sigma_1(\rho,\xi)
	&=& \log\left(|\xi|^2 + \tilde{V}_{\rho}(\xi) + \rho^{-1}\right)\int_0^{\kappa(\rho,\xi)} \e^{-w} \intd w - \int_0^{\kappa(\rho,\xi)} \log\left( \tilde{V}_{\rho}(\xi) + \rho^{-1} \right) \e^{-w} \intd w \nonumber\\
	&& { } + \int_{\kappa(\rho,\xi)}^\infty \log\left(1 + \frac{|\xi|^2}{\tilde{V}_{\rho}(\xi) + \rho^{-1}}\right) \e^{-w} \intd w   \label{Sigma_1} \\
	\Sigma_2(\rho,\xi)	&=& \int_0^{\kappa(\rho,\xi)} \log\left(\tilde{V}_{\rho}(\xi) + \rho^{-1}\right) \e^{-w} \intd w - \int_0^{\kappa(\rho,\xi)} \log\left(\tilde{\Phi}_{\rho}(\xi)w + \rho^{-1}\right) \e^{-w} \intd w \nonumber\\
	&& { } + \int_{\kappa(\rho,\xi)}^\infty \log\left(\frac{\tilde{V}_{\rho}(\xi) + \rho^{-1}}{\tilde{\Phi}_{\rho}(\xi)w + \rho^{-1}}\right) \e^{-w} \intd w   \label{Sigma_2} \\
	\Sigma_3(\rho,\xi)
	&=& \int_0^{\kappa(\rho,\xi)} \frac{|\xi|^2}{(1-w)\tilde{V}_{\rho}(\xi) + |\xi|^2} \log\left( \tilde{V}_{\rho}(\xi)w + \rho^{-1} \right) \e^{-w} \intd w \nonumber\\
	&& { } - \log\left( |\xi|^2 + \tilde{V}_{\rho}(\xi) + \rho^{-1} \right) \int_0^{\kappa(\rho,\xi)} \frac{|\xi|^2}{(1-w)\tilde{V}_{\rho}(\xi) + |\xi|^2} \e^{-w} \intd w.   \label{Sigma_3} 
\end{IEEEeqnarray}
\end{subequations}
Upon reordering terms in \eqref{Sigma_1}--\eqref{Sigma_3}, the upper bound \eqref{Sigma_upper_bound_1} can be further rewritten as
\begin{align}   \label{Sigma_upper_bound_2}
	\Sigma(\rho,\xi)
	\leq \sum_{i=1}^5 J_i(\rho,\xi)
\end{align}
with the five terms
\begin{subequations}
\begin{IEEEeqnarray}{lCl}
	J_1(\rho,\xi)
	&\triangleq & \int_0^{\kappa(\rho,\xi)} \log\left( \frac{\tilde{V}_{\rho}(\xi)w + \rho^{-1}}{\tilde{\Phi}_{\rho}(\xi)w + \rho^{-1}} \right) \e^{-w} \intd w   \label{J_1} \\
	J_2(\rho,\xi)
	& \triangleq & -\int_0^{\kappa(\rho,\xi)} \frac{(1-w)\tilde{V}_{\rho}(\xi)}{(1-w)\tilde{V}_{\rho}(\xi) + |\xi|^2} \log\bigl(\tilde{V}_{\rho}(\xi)w + \rho^{-1}\bigr) \e^{-w} \intd w   \label{J_2} \\
	J_3(\rho,\xi)
	& \triangleq & \log\bigl(|\xi|^2 + \tilde{V}_{\rho}(\xi) + \rho^{-1}\bigr) \int_0^{\kappa(\rho,\xi)} \frac{(1-w)\tilde{V}_{\rho}(\xi)}{(1-w)\tilde{V}_{\rho}(\xi) + |\xi|^2} \e^{-w} \intd w   \label{J_3} \\
	J_4(\rho,\xi) 
	& \triangleq & \int_{\kappa(\rho,\xi)}^\infty \log\left(\frac{\tilde{V}_{\rho}(\xi) + \rho^{-1}}{\tilde{\Phi}_{\rho}(\xi)w+\rho^{-1}}\right) \e^{-w} \intd w   \label{J_4} \\
	J_5(\rho,\xi)
	& \triangleq & \log\left(1 + \frac{|\xi|^2}{\tilde{V}_{\rho}(\xi) + \rho^{-1}}\right) e^{-\kappa(\rho,\xi)}.   \label{J_5}
\end{IEEEeqnarray}
\end{subequations}
Here, the term $J_1(\rho,\xi)$ is the second term of \eqref{Sigma_2} to which we add $\int_0^{\kappa(\rho,\xi)} \log\bigl(\tilde{V}_{\rho}(\xi)w + \rho^{-1}\bigr) \e^{-w} \intd w$; the term $J_2(\rho,\xi)$ is the first term in \eqref{Sigma_3} from which we subtract $\int_0^{\kappa(\rho,\xi)} \log\bigl(\tilde{V}_{\rho}(\xi)w + \rho^{-1}\bigr) \e^{-w} \intd w$; the term $J_3(\rho,\xi)$ follows from adding the first term in \eqref{Sigma_1} to the second term in \eqref{Sigma_3}; the term $J_4(\rho,\xi)$ is the third term in \eqref{Sigma_2}; the term $J_5(\rho,\xi)$ is the third term in \eqref{Sigma_1}. The second term of \eqref{Sigma_1} and the first term of \eqref{Sigma_2} cancel out.

We proceed by showing that, for every $\xi_0>0$,
\begin{subequations}
\begin{IEEEeqnarray}{rCll}
\varlimsup_{\rho\to\infty}\Exp\bigl[J_1(\rho,\hat{H}_{\rho})\operatorname{I}\bigl\{|\hat{H}_{\rho}| > \xi_0\bigr\}\bigr] & \leq & \log(M) \quad & \label{eq:doby_Sigma2_J1}\\
\varlimsup_{\rho\to\infty} \Exp\bigl[J_i(\rho,\hat{H}_{\rho})\operatorname{I}\bigl\{|\hat{H}_{\rho}| > \xi_0\bigr\}\bigr] & \leq & 0, \quad & i=2,3,4,5. \label{eq:doby_Sigma2_Ji}
\end{IEEEeqnarray}
\end{subequations}
The claim \eqref{eq:doby_Sigma2} then follows by combining \eqref{eq:doby_Sigma2_J1} and \eqref{eq:doby_Sigma2_Ji} with \eqref{Sigma_upper_bound_2} and by letting $\xi_0$ tend to zero from above. The following lemma will be useful.
\begin{mylem}   \label{lem:continuity_at_0}
Consider the family of random variables $\Upsilon_{\rho}$ parametrized by $\rho>0$ taking values on $(0,\eta)$ and satisfying $\lim_{\rho \rightarrow \infty} \Exp[\Upsilon_{\rho}] = 0$, where $\eta$ belongs to the extended positive reals, i.e., $\eta \in (0,\infty]$. Let $f(\cdot)$ be a continuous bounded function on the interval $(0,\eta)$ with limit $\lim_{t \downarrow 0} f(t) = f_0$. Then
\begin{align}
	\lim_{\rho \rightarrow \infty} \Exp\bigl[ f(\Upsilon_{\rho}) \bigr] = f_0.
\end{align}
\end{mylem}
\begin{IEEEproof}
See Appendix~\ref{app:lem:proof:continuity_at_0}.
\end{IEEEproof}

\subsubsection{Limit related to $J_1(\rho,\xi)$}

Noting that $\tilde{\Phi}_{\rho}(\xi) \leq \tilde{V}_{\rho}(\xi)$, we have that
\begin{equation*}
w\mapsto \frac{\tilde{V}_{\rho}(\xi)w + \rho^{-1}}{\tilde{\Phi}_{\rho}(\xi)w + \rho^{-1}}
\end{equation*}
is monotonically increasing in $w$. Consequently, $J_1(\rho,\xi)$ is upper-bounded by
\begin{align}   \label{J_1_upper_bound}
	J_1(\rho,\xi)
	&\leq \int_0^{\kappa(\rho,\xi)} \log\frac{\tilde{V}_{\rho}(\xi)}{\tilde{\Phi}_{\rho}(\xi)} \e^{-w} \intd w \nonumber\\
	&\leq \left[1 - e^{-\frac{\xi_0^2}{\sqrt{\tilde{V}_{\rho}(\xi)+\rho^{-1}}}} \right] \log\left(\sup_{\xi\in\mathbb{C}}\frac{\tilde{V}_{\rho}(\xi)}{\tilde{\Phi}_{\rho}(\xi)}\right), \quad \bigl(\rho>0,\,|\xi|>\xi_0\bigr)
\end{align}
where in the last step we have used \eqref{kappa}.
Setting $\xi$ to $\hat{H}_{\rho}$, averaging \eqref{J_1_upper_bound} over $\hat{H}_{\rho}$, and upper-bounding
\begin{equation}
\operatorname{I}\bigl\{|\hat{H}_{\rho}| > \xi_0\bigr\} \leq 1 \label{eq:doby_UB_I}
\end{equation}
we obtain
\begin{IEEEeqnarray}{lCl}
\Exp\bigl[J_1(\rho,\hat{H}_{\rho})\operatorname{I}\bigl\{|\hat{H}_{\rho}| > \xi_0\bigr\}\bigr] & \leq & \left(1-\Exp\left[e^{-\frac{\xi_0^2}{\sqrt{\tilde{V}_{\rho}(\hat{H}_{\rho})+\rho^{-1}}}}\right]\right)\log\left(\sup_{\xi\in\mathbb{C}}\frac{\tilde{V}_{\rho}(\xi)}{\tilde{\Phi}_{\rho}(\xi)}\right), \quad \rho>0.  \IEEEeqnarraynumspace\label{eq:J1_avg}
\end{IEEEeqnarray}
Noting that the function $t \mapsto \exp\left(-\xi_0^2/\sqrt{t}\right)$
is continuous and bounded on $(0,\infty)$ and vanishes as $t$ tends to zero, it follows from \eqref{eq:thm_Vtilde} and Lemma~\ref{lem:continuity_at_0} that
\begin{equation}
\label{eq:J1_UB}
\lim_{\rho\to\infty}\Exp\left[\exp\left(-\frac{\xi_0^2}{\sqrt{\tilde{V}_{\rho}(\hat{H}_{\rho})+\rho^{-1}}}\right)\right] =0.
\end{equation}
We further have by \eqref{technical_condition} as well as the continuity and monotonicity of $x\mapsto \log(x)$ that
\begin{equation}
\varlimsup_{\rho\to\infty} \log\left(\sup_{\xi\in\mathbb{C}}\frac{\tilde{V}_{\rho}(\xi)}{\tilde{\Phi}_{\rho}(\xi)}\right) \leq \log(M). \label{eq:doby_J1_M}
\end{equation}
Combining \eqref{eq:J1_UB} and \eqref{eq:doby_J1_M} with \eqref{eq:J1_avg} proves \eqref{eq:doby_Sigma2_J1}.

\subsubsection{Limit related to $J_2(\rho,\xi)$}

To prove \eqref{eq:doby_Sigma2_Ji} for $i=2$, first note that $0<\xi_0<1$ implies that, for sufficiently large $\rho$, we have
\begin{equation}
\tilde{V}_{\rho}(\xi)w+\rho^{-1} \leq 1, \quad 0\leq w \leq \kappa(\rho,\xi)
\end{equation}
and
\begin{equation}
(1-w)\tilde{V}_{\rho}(\xi) \geq -|\xi|^2, \quad 0\leq w \leq \kappa(\rho,\xi).
\end{equation}
Further note that $t\mapsto t/(t+|\xi|^2)$ is monotonically increasing on $(-|\xi|^2,\infty)$.
Consequently, for sufficiently large $\rho$, \eqref{J_2} is upper-bounded by
\begin{IEEEeqnarray}{lCl}
J_2(\rho,\xi) & \leq & - \frac{\tilde{V}_{\rho}(\xi)}{\tilde{V}_{\rho}(\xi) + |\xi|^2}\int_0^{\kappa(\rho,\xi)} \log\bigl(\tilde{V}_{\rho}(\xi)w + \rho^{-1}\bigr) \e^{-w} \intd w \nonumber\\
& \leq & \frac{\tilde{V}_{\rho}(\xi)}{\tilde{V}_{\rho}(\xi) + |\xi|^2}\left[\left(1-e^{-\kappa(\xi,\rho)}\right)\log\frac{1}{\tilde{V}_{\rho}(\xi)}+\int_0^{\kappa(\rho,\xi)}\left|\log(w)\right| e^{-w}\intd w\right] \label{doby_J4_1}
\end{IEEEeqnarray}
where the second inequality follows by lower-bounding $\log\bigl(\tilde{V}_{\rho}(\xi)w+\rho^{-1}\bigr)\geq \log\bigl(\tilde{V}_{\rho}(\xi)\bigr)+\log(w)$ and from the triangle inequality.

By using that the exponential function is nonnegative, by upper-bounding the integral by integrating to infinity, and by using that $|\xi| > \xi_0$, we can further upper-bound \eqref{doby_J4_1}, for sufficiently large $\rho$, by
\begin{equation}
J_2(\rho,\xi) \leq  \frac{\tilde{V}_{\rho}(\xi)}{\tilde{V}_{\rho}(\xi) + \xi_0^2}\left[\log\frac{1}{\tilde{V}_{\rho}(\xi)}+K\right] \label{doby_J4_2}
\end{equation}
where we define
\begin{align}   \label{K}
	K
	& \triangleq \int_0^{\infty}\left|\log(w)\right| \e^{-w} \intd w = \gamma - 2\expint(-1)
\end{align}
and where $\operatorname{Ei}(\cdot)$ denotes the exponential integral function, i.e.,
\begin{equation}   \label{def:exp_integral}
\operatorname{Ei}(-x) \triangleq -\int_x^{\infty}\frac{e^{-u}}{u} \textnormal{d} u.
\end{equation}
Noting that the RHS of \eqref{doby_J4_2} is a continuous and bounded function of $0<\tilde{V}_{\rho}(\xi)\leq 1$ that vanishes as $\tilde{V}_{\rho}(\xi)$ tends to zero, it follows from \eqref{doby_J4_2}, \eqref{eq:doby_UB_I}, \eqref{eq:thm_Vtilde}, and Lemma~\ref{lem:continuity_at_0} that
\begin{equation}
\varlimsup_{\rho\to\infty} \Exp\bigl[J_2(\rho,\hat{H}_{\rho})\operatorname{I}\bigl\{|\hat{H}_{\rho}| > \xi_0\bigr\}\bigr] \leq \varlimsup_{\rho\to\infty} \Exp\left[\frac{\tilde{V}_{\rho}(\hat{H}_{\rho})}{\tilde{V}_{\rho}(\hat{H}_{\rho}) + \xi_0^2}\left(\log\frac{1}{\tilde{V}_{\rho}(\hat{H}_{\rho})}+K\right)\right] \leq 0
\end{equation}
thus proving \eqref{eq:doby_Sigma2_Ji} for $i=2$.

\subsubsection{Limit related to $J_3(\rho,\xi)$}

To prove \eqref{eq:doby_Sigma2_Ji} for $i=3$, we shall prove the stronger statement
\begin{equation}
\lim_{\rho\to\infty} \Exp\left[\bigl|J_3(\rho,\hat{H}_{\rho})\bigr|\operatorname{I}\bigl\{|\hat{H}_{\rho}| > \xi_0\bigr\}\right] = 0. \label{doby_J3_stronger}
\end{equation}
To this end, note that by the triangle inequality
\begin{align}   \label{integrand_framed}
	\left|\frac{(1-w)\tilde{V}_{\rho}(\xi)}{(1-w)\tilde{V}_{\rho}(\xi) + |\xi|^2}\right| \leq \frac{\bigl(1+\kappa(\rho,\xi)\bigr)\tilde{V}_{\rho}(\xi)}{\bigl(1-\kappa(\rho,\xi)\bigr)\tilde{V}_{\rho}(\xi)+\xi_0^2}, \quad 0\leq w\leq \kappa(\rho,\xi).
\end{align}
In \eqref{integrand_framed} we have used that, for $0 \leq w \leq \kappa(\rho,\xi)$ and $|\xi|\geq\xi_0$, the denominator is lower-bounded by $\bigl(1-\kappa(\rho,\xi)\bigr)\tilde{V}_{\rho}(\xi)+\xi_0^2 > 0$. It follows from \eqref{integrand_framed} and the triangle inequality that the absolute value of the integral in \eqref{J_3} is upper-bounded by
\begin{IEEEeqnarray}{lCl}
\left|\int_0^{\kappa(\rho,\xi)} \frac{(1-w)\tilde{V}_{\rho}(\xi)}{(1-w)\tilde{V}_{\rho}(\xi) + |\xi|^2} \e^{-w} \intd w\right| & \leq & \left( 1 - \e^{-\kappa(\rho,\xi)} \right) \frac{\bigl(1+\kappa(\rho,\xi)\bigr)\tilde{V}_{\rho}(\xi)}{\bigl(1-\kappa(\rho,\xi)\bigr)\tilde{V}_{\rho}(\xi)+\xi_0^2}. \label{integrand_framed_2}
\end{IEEEeqnarray}
Consequently,
\begin{IEEEeqnarray}{lCl}
\bigl|J_3(\rho,\xi)\bigr| & \leq & \left( 1 - \e^{-\kappa(\rho,\xi)} \right) \frac{\bigl(1+\kappa(\rho,\xi)\bigr)\tilde{V}_{\rho}(\xi)}{\bigl(1-\kappa(\rho,\xi)\bigr)\tilde{V}_{\rho}(\xi)+\xi_0^2} \left|\log\bigl(|\xi|^2+\tilde{V}_{\rho}(\xi) + \rho^{-1}\bigr)\right| \nonumber\\
& \leq & \frac{\bigl(1+\kappa(\rho,\xi)\bigr)\bigl(\tilde{V}_{\rho}(\xi)+\rho^{-1}\bigr)}{\xi_0^2-\bigl(\kappa(\rho,\xi)-1\bigr)^+ \bigl(\tilde{V}_{\rho}(\xi)+\rho^{-1}\bigr)} \left|\log\bigl(|\xi|^2 + \tilde{V}_{\rho}(\xi) + \rho^{-1}\bigr)\right|, \quad \bigl(\rho>0,\,|\xi|\geq \xi_0\bigr) \IEEEeqnarraynumspace\label{J_3_upper_bound}
\end{IEEEeqnarray}
where we define $(a)^+\triangleq \max(a,0)$.\footnote{The condition $\xi_0< 1$ ensures that the denominator remains positive.} Here the last step follows by upper-bounding $\tilde{V}_{\rho}(\xi)\leq \tilde{V}_{\rho}(\xi)+\rho^{-1}$ and by lower-bounding $\bigl(1-\kappa(\rho,\xi)\bigr)\tilde{V}_{\rho}(\xi)\geq -\bigl(\kappa(\rho,\xi)-1\bigr)^+ \bigl(\tilde{V}_{\rho}(\xi)+\rho^{-1}\bigr)$ and $e^{-\kappa(\rho,\xi)}\geq 0$.

Using the definition \eqref{kappa} of $\kappa(\rho,\xi)$, and defining $\Upsilon_{\rho}(\xi)\triangleq \tilde{V}_{\rho}(\xi)+\rho^{-1}$, the RHS of \eqref{J_3_upper_bound} reads as
\begin{equation}
\frac{\Upsilon_{\rho}(\xi)+\xi_0^2 \sqrt{\Upsilon_{\rho}(\xi)}}{\xi_0^2-\bigl(\xi_0^2-\sqrt{\Upsilon_{\rho}(\xi)}\bigr)^+ \sqrt{\Upsilon_{\rho}(\xi)}}\left|\log\bigl(|\xi|^2+\Upsilon_{\rho}(\xi)\bigr)\right|. \label{doby_J3_UB_1}
\end{equation}
Since $\tilde{V}_{\rho}(\xi)\leq 1$ and $x\mapsto \log(x)$ is a monotonically increasing function, we have
\begin{equation}
\log\bigl(\Upsilon_{\rho}(\xi)\bigr) \leq \log\bigl(|\xi|^2+\Upsilon_{\rho}(\xi)\bigr) \leq  \log\left(1+\rho^{-1}+|\xi|^2\right).
\end{equation}
The absolute value of the logarithm on the RHS of \eqref{doby_J3_UB_1} is thus upper-bounded by
\begin{equation}
\left|\log\bigl(|\xi|^2+\Upsilon_{\rho}(\xi)\bigr)\right| \leq \left|\log\bigl(\Upsilon_{\rho}(\xi)\bigr)\right| + \log\left(1+\rho^{-1}+|\xi|^2\right). \label{doby_J3_max}
\end{equation}
By noting that
\begin{equation}
\Upsilon_{\rho}(\xi) \mapsto \frac{\Upsilon_{\rho}(\xi)+\xi_0^2 \sqrt{\Upsilon_{\rho}(\xi)}}{\xi_0^2-\bigl(\xi_0^2-\sqrt{\Upsilon_{\rho}(\xi)}\bigr)^+ \sqrt{\Upsilon_{\rho}(\xi)}}\left|\log\bigl(\Upsilon_{\rho}(\xi)\bigr)\right|
\end{equation}
is a continuous and bounded function of $0<\Upsilon_{\rho}(\xi)\leq 1+\rho^{-1}$ that vanishes as $\Upsilon_{\rho}(\xi)$ tends to zero, we obtain from \eqref{eq:thm_Vtilde}, \eqref{eq:doby_UB_I}, and Lemma~\ref{lem:continuity_at_0} that
\begin{IEEEeqnarray}{lCl}
\IEEEeqnarraymulticol{3}{l}{\lim_{\rho\to\infty}\Exp\left[\frac{\Upsilon_{\rho}(\hat{H}_{\rho})+\xi_0^2 \sqrt{\Upsilon_{\rho}(\hat{H}_{\rho})}}{\xi_0^2-\bigl(\xi_0^2-\sqrt{\Upsilon_{\rho}(\hat{H}_{\rho})}\bigr)^+ \sqrt{\Upsilon_{\rho}(\hat{H}_{\rho})}}\bigl|\log\bigl(\Upsilon_{\rho}(\hat{H}_{\rho})\bigr)\bigr|\operatorname{I}\bigl\{|\hat{H}_{\rho}| > \xi_0\bigr\}\right]} \nonumber\\
\quad & \leq & 
\lim_{\rho\rightarrow\infty}\Exp\left[\frac{\Upsilon_{\rho}(\hat{H}_{\rho})+\xi_0^2 \sqrt{\Upsilon_{\rho}(\hat{H}_{\rho})}}{\xi_0^2-\bigl(\xi_0^2-\sqrt{\Upsilon_{\rho}(\hat{H}_{\rho})}\bigr)^+ \sqrt{\Upsilon_{\rho}(\hat{H}_{\rho})}}\bigl|\log\bigl(\Upsilon_{\rho}(\hat{H}_{\rho})\bigr)\bigr|\right] \nonumber\\
& = & 0. \label{doby_J3_yes}
\end{IEEEeqnarray}
Furthermore, \eqref{eq:doby_UB_I} together with the Cauchy-Schwarz inequality yields
\begin{IEEEeqnarray}{lCl}
\IEEEeqnarraymulticol{3}{l}{\Exp\left[\frac{\Upsilon_{\rho}(\hat{H}_{\rho})+\xi_0^2 \sqrt{\Upsilon_{\rho}(\hat{H}_{\rho})}}{\xi_0^2-\bigl(\xi_0^2-\sqrt{\Upsilon_{\rho}(\hat{H}_{\rho})}\bigr)^+ \sqrt{\Upsilon_{\rho}(\hat{H}_{\rho})}}\log\left(1+\rho^{-1}+|\hat{H}_{\rho}|^2\right)\operatorname{I}\bigl\{|\hat{H}_{\rho}| > \xi_0\bigr\}\right]}\nonumber\\
\qquad & \leq & \Exp\left[\frac{\Upsilon_{\rho}(\hat{H}_{\rho})+\xi_0^2 \sqrt{\Upsilon_{\rho}(\hat{H}_{\rho})}}{\xi_0^2-\bigl(\xi_0^2-\sqrt{\Upsilon_{\rho}(\hat{H}_{\rho})}\bigr)^+ \sqrt{\Upsilon_{\rho}(\hat{H}_{\rho})}}\log\left(1+\rho^{-1}+|\hat{H}_{\rho}|^2\right)\right] \nonumber\\
\qquad & \leq & \sqrt{\Exp\left[\left(\frac{\Upsilon_{\rho}(\hat{H}_{\rho})+\xi_0^2 \sqrt{\Upsilon_{\rho}(\hat{H}_{\rho})}}{\xi_0^2-\bigl(\xi_0^2-\sqrt{\Upsilon_{\rho}(\hat{H}_{\rho})}\bigr)^+ \sqrt{\Upsilon_{\rho}(\hat{H}_{\rho})}}\right)^2\right]}\sqrt{\Exp\left[\log^2\left(1+\rho^{-1}+|\hat{H}_{\rho}|^2\right)\right]}.\label{eq:doby_J3_bla}
\end{IEEEeqnarray}
Note that the term inside the first expected value is a continuous and bounded function of $0<\Upsilon_{\rho}(\hat{H}_{\rho})\leq 1+\rho^{-1}$ that vanishes as $\Upsilon_{\rho}(\hat{H}_{\rho})$ tends to zero, so it follows from \eqref{eq:thm_Vtilde} and Lemma~\ref{lem:continuity_at_0} that the first expected value on the RHS of \eqref{eq:doby_J3_bla} vanishes as $\rho$ tends to infinity. We further show in Appendix~\ref{sub:doby_bounded_log2} that
\begin{equation}
\label{eq:doby_bounded_log2}
\varlimsup_{\rho\to\infty} \Exp\left[\log^2\left(1+\rho^{-1}+|\hat{H}_{\rho}|^2\right)\right] < \infty.
\end{equation}
The above arguments combine to demonstrate that
\begin{equation}
\lim_{\rho\to\infty}\Exp\left[\frac{\Upsilon_{\rho}(\hat{H}_{\rho})+\xi_0^2 \sqrt{\Upsilon_{\rho}(\hat{H}_{\rho})}}{\xi_0^2-\bigl(\xi_0^2-\sqrt{\Upsilon_{\rho}(\hat{H}_{\rho})}\bigr)^+ \sqrt{\Upsilon_{\rho}(\hat{H}_{\rho})}}\log\left(1+\rho^{-1}+|\hat{H}_{\rho}|^2\right)\operatorname{I}\bigl\{|\hat{H}_{\rho}| > \xi_0\bigr\}\right] = 0. \label{doby_J3_almost}
\end{equation}
Combining \eqref{doby_J3_almost}, \eqref{doby_J3_yes}, \eqref{doby_J3_max}, and \eqref{J_3_upper_bound} proves \eqref{doby_J3_stronger}.

\subsubsection{Limit related to $J_4(\rho,\xi)$}

To upper-bound $J_4(\rho,\xi)$, we use that, for $w\geq \kappa(\rho,\xi)$,
\begin{IEEEeqnarray}{lCl}
\frac{\tilde{V}_{\rho}(\xi) + \rho^{-1}}{\tilde{\Phi}_{\rho}(\xi)w+\rho^{-1}} & = & \frac{\tilde{V}_{\rho}(\xi)}{\tilde{\Phi}_{\rho}(\xi)w+\rho^{-1}} + \frac{\rho^{-1}}{\tilde{\Phi}_{\rho}(\xi)w+\rho^{-1}} \nonumber\\
& \leq & \frac{\tilde{V}_{\rho}(\xi)}{\tilde{\Phi}_{\rho}(\xi)\kappa(\rho,\xi)} + 1 \nonumber\\
& \leq & \sup_{\xi\in\mathbb{C}}\biggl\{\frac{\tilde{V}_{\rho}(\xi)}{\tilde{\Phi}_{\rho}(\xi)}\biggr\} \frac{\sqrt{1+\rho^{-1}}}{\xi_0^2} + 1 \label{eq:J2_1}
\end{IEEEeqnarray}
where the first inequality follows by lower-bounding $\rho^{-1}\geq 0$ and $w\geq\kappa(\rho,\xi)$ in the denominator of the first fraction and by lower-bounding $\tilde{\Phi}_{\rho}(\xi)w\geq 0$ in the denominator of the second fraction; and where the second inequality follows by lower-bounding $\kappa(\rho,\xi)\geq \xi_0^2/\sqrt{1+\rho^{-1}}$ using $\tilde{V}_{\rho}(\xi)\leq 1$ and by maximizing over $\xi$. Combining \eqref{eq:J2_1} with \eqref{J_4} yields
\begin{align}
J_4(\rho,\xi) & \leq \log\left(1+\sup_{\xi\in\mathbb{C}}\biggl\{\frac{\tilde{V}_{\rho}(\xi)}{\tilde{\Phi}_{\rho}(\xi)}\biggr\}\frac{\sqrt{1+\rho^{-1}}}{\xi_0^2}\right) \int_{\kappa(\rho,\xi)}^{\infty} e^{-w} \intd w \nonumber\\
& = \exp\left(-\frac{\xi_0^2}{\sqrt{\tilde{V}_{\rho}(\xi)+\rho^{-1}}}\right)\log\left(1+\sup_{\xi\in\mathbb{C}}\biggl\{\frac{\tilde{V}_{\rho}(\xi)}{\tilde{\Phi}_{\rho}(\xi)}\biggr\}\frac{\sqrt{1+\rho^{-1}}}{\xi_0^2}\right). \label{eq:J2_2}
\end{align}
Setting $\xi$ to $\hat{H}_{\rho}$, averaging \eqref{eq:J2_2} over $\hat{H}_{\rho}$, and using \eqref{eq:doby_UB_I}, we obtain
\begin{IEEEeqnarray}{lCl}
\IEEEeqnarraymulticol{3}{l}{\Exp\bigl[J_4(\rho,\hat{H}_{\rho})\operatorname{I}\bigl\{|\hat{H}_{\rho}| > \xi_0\bigr\}\bigr]}\nonumber\\
\quad  & \leq & \Exp\left[ \exp\left(-\frac{\xi_0^2}{\sqrt{\tilde{V}_{\rho}(\hat{H}_{\rho})+\rho^{-1}}}\right)\right]\log\left(1+\sup_{\xi\in\mathbb{C}}\biggl\{\frac{\tilde{V}_{\rho}(\xi)}{\tilde{\Phi}_{\rho}(\xi)}\biggr\}\frac{\sqrt{1+\rho^{-1}}}{\xi_0^2}\right). \IEEEeqnarraynumspace \label{eq:J2_3}
\end{IEEEeqnarray}
Since, by \eqref{technical_condition}, the term inside the logarithm is bounded for sufficiently large $\rho$, \eqref{eq:doby_Sigma2_Ji} for $i=4$ follows by combining \eqref{eq:J2_3} with \eqref{eq:J1_UB}.

\subsubsection{Limit related to $J_5(\rho,\xi)$}
Using \eqref{kappa} and defining $\Upsilon_{\rho}(\xi)\triangleq \tilde{V}_{\rho}(\xi)+\rho^{-1}$, the term $J_5(\rho,\xi)$ reads as
\begin{equation}
	J_5(\rho,\xi) = e^{-\frac{\xi_0^2}{\sqrt{\Upsilon_{\rho}(\xi)}}}\log\left(1+\frac{|\xi|^2}{\Upsilon_{\rho}(\xi)}\right).\end{equation}
Since $\tilde{V}_{\rho}(\xi)\leq 1$ and $x\mapsto \log(x)$ is a monotonically increasing function, this can be upper-bounded as
\begin{equation}
\label{eq:doby_J_5_first_bound}
J_5(\rho,\xi) \leq e^{-\frac{\xi_0^2}{\sqrt{\Upsilon_{\rho}(\xi)}}}\log\left(1+\rho^{-1}+|\xi|^2\right) + e^{-\frac{\xi_0^2}{\sqrt{\Upsilon_{\rho}(\xi)}}}\bigl|\log\bigl(\Upsilon_{\rho}(\xi)\bigr)\bigr|.
\end{equation}
We next note that the function $t\mapsto e^{-\xi_0^2/\sqrt{t}}|\log(t)|$ is continuous and bounded on $(0,1+\rho^{-1})$ and tends to zero as $t$ tends to zero. Consequently, \eqref{eq:doby_UB_I} and Lemma~\ref{lem:continuity_at_0} yield
\begin{equation}
\lim_{\rho\to\infty} \Exp\left[e^{-\frac{\xi_0^2}{\sqrt{\Upsilon_{\rho}(\hat{H}_{\rho})}}}\bigl|\log\bigl(\Upsilon_{\rho}(\hat{H}_{\rho})\bigr)\bigr|\cdot\operatorname{I}\bigl\{|\hat{H}_{\rho}| > \xi_0\bigr\}\right] =0. \label{eq:doby_J_5_upper_bound_bla}
\end{equation}
Furthermore, by \eqref{eq:doby_UB_I} and the Cauchy-Schwarz inequality we have
\begin{IEEEeqnarray}{rCl}   \label{J_5_upper_bound}
	\Exp\left[e^{-\frac{\xi_0^2}{\sqrt{\Upsilon_{\rho}(\hat{H}_{\rho})}}}\log\left(1+\rho^{-1}+|\hat{H}_{\rho}|^2\right)\operatorname{I}\bigl\{|\hat{H}_{\rho}| > \xi_0\bigr\}\right]
	&\leq& \Exp\left[e^{-\frac{\xi_0^2}{\sqrt{\Upsilon_{\rho}(\hat{H}_{\rho})}}}\log\left(1+\rho^{-1}+|\hat{H}_{\rho}|^2\right)\right] \nonumber\\
	&\leq& \sqrt{\Exp\left[e^{-\frac{2\xi_0^2}{\sqrt{\Upsilon_{\rho}(\hat{H}_{\rho})}}}\right]} \sqrt{\Exp\left[ \log^2\left(1+\rho^{-1}+|\hat{H}_{\rho}|^2\right)\right]}. \IEEEeqnarraynumspace
\end{IEEEeqnarray}
Since the function $t\mapsto \exp\left(-2\xi_0^2/\sqrt{t}\right)$ is continuous and bounded on $(0,\infty)$ and vanishes as $t$ tends to zero, it follows from \eqref{technical_condition} and Lemma~\ref{lem:continuity_at_0}  that the first expected value on the RHS of \eqref{J_5_upper_bound} vanishes as $\rho$ tends to infinity. Furthermore, by \eqref{eq:doby_bounded_log2}, the second expected value on the RHS of \eqref{J_5_upper_bound} is bounded for sufficiently small $\rho$. The above arguments combine to demonstrate that
\begin{equation}
\lim_{\rho\to\infty} \Exp\left[e^{-\frac{\xi_0^2}{\sqrt{\Upsilon_{\rho}(\hat{H}_{\rho})}}}\log\left(1+\rho^{-1}+|\hat{H}_{\rho}|^2\right)\operatorname{I}\bigl\{|\hat{H}_{\rho}| > \xi_0\bigr\}\right] = 0
\end{equation}
which together with \eqref{eq:doby_J_5_first_bound} and \eqref{eq:doby_J_5_upper_bound_bla} proves \eqref{eq:doby_Sigma2_Ji} for $i=5$.

\subsection{Proof of Lemma~\ref{lem:g}}   \label{app:lem:proof:g}
We first note that, by specializing Theorem~\ref{cor:layer_monotonicity} to the case where $\hat{H}=\xi$ with probability one, it follows that
\begin{equation}
\label{eq:doby_lemm8_the_latter}
R_{\textnormal{M}}(\rho,\xi) \leq \Exp\bigl[R^{\star}(\rho,W,\xi)\bigr], \quad \bigl(\rho>0,\,\xi\in\mathbb{C}\bigr).
\end{equation}
Combining \eqref{eq:doby_lemm8_the_latter} with \eqref{Delta_three_arguments} and \eqref{Sigma_two_arguments}, we obtain
\begin{IEEEeqnarray}{lCl}
	\Sigma(\rho,\xi) & \leq & \Exp\bigl[\Delta(\rho,W,\xi)\bigr] \nonumber\\
	&= & \log\left(\frac{\tilde{V}_{\rho}(\xi) + \rho^{-1}}{\tilde{\Phi}_{\rho}(\xi)}\right) - \Exp\left[\log\left(W + \frac{1}{\rho \tilde{\Phi}_{\rho}(\xi)}\right)\right]. \label{eq:appF_1}
\end{IEEEeqnarray}
The expected value on the RHS of \eqref{eq:appF_1} can be evaluated as \cite[(4.337),~p.~568]{GrRy94}
\begin{equation}
\Exp\left[\log\left(W + \frac{1}{\rho \tilde{\Phi}_{\rho}(\xi)}\right)\right] = \log\left(\frac{1}{\rho \tilde{\Phi}_{\rho}(\xi)}\right) - e^{\frac{1}{\rho\tilde{\Phi}_{\rho}(\xi)}}\operatorname{Ei}\left(-\frac{1}{\rho \tilde{\Phi}_{\rho}(\xi)}\right)
\end{equation}
where $\operatorname{Ei}(\cdot)$ denotes the exponential integral as defined in \eqref{def:exp_integral}. This yields
\begin{align}
\Exp\bigl[\Delta(\rho,W,\xi)\bigr] & = \log\bigl(1+\rho\tilde{V}_{\rho}(\xi)\bigr) + e^{\frac{1}{\rho\tilde{\Phi}_{\rho}(\xi)}}\operatorname{Ei}\left(-\frac{1}{\rho \tilde{\Phi}_{\rho}(\xi)}\right) \nonumber\\
& = \log\left(1+\rho\tilde{\Phi}_{\rho}(\xi)\frac{\tilde{V}_{\rho}(\xi)}{\tilde{\Phi}_{\rho}(\xi)}\right) + e^{\frac{1}{\rho\tilde{\Phi}_{\rho}(\xi)}}\operatorname{Ei}\left(-\frac{1}{\rho \tilde{\Phi}_{\rho}(\xi)}\right) \nonumber\\
& \leq \log\left(1+\rho\tilde{\Phi}_{\rho}(\xi)\frac{\tilde{V}_{\rho}(\xi)}{\tilde{\Phi}_{\rho}(\xi)}\right) + \operatorname{Ei}\left(-\frac{1}{\rho \tilde{\Phi}_{\rho}(\xi)}\right) \nonumber\\
& = g\left(\rho\tilde{\Phi}_{\rho}(\xi);\frac{\tilde{V}_{\rho}(\xi)}{\tilde{\Phi}_{\rho}(\xi)}\right) \label{eq:appF_g}
\end{align}
where we define
\begin{equation}
g(t;a) \triangleq \log(1+at) + \operatorname{Ei}\left(-\frac{1}{t}\right).
\end{equation}
The inequality in \eqref{eq:appF_g} follows because $\operatorname{Ei}(-x)$ is negative for $x>0$ and $e^{x}\geq 1$, $x\geq 0$.

For a fixed $a$, the function $t\mapsto g(t;a)$ satisfies \cite[Section~VI-A]{LaMo03}\footnote{The function $g(\cdot;\cdot)$ corresponds to $g_0(\cdot)$ in \cite[Equation~(210)]{LaMo03} via $g(t;a)=\log(a)+\log\left(1+\frac{1}{at}\right)-g_0\left(\frac{1}{t}\right)$. The result \eqref{eq:appF_glim} follows by noting that $g_0(0)=-\gamma$; cf.\ \cite[Equations~(212) and (213)]{LaMo03}.}
\begin{equation}
\lim_{t\to\infty} g(t;a) =\gamma + \log(a). \label{eq:appF_glim}
\end{equation}
We next show that, for every $a\geq 1$, the function $t\mapsto g(t;a)$ is monotonically increasing. Indeed, using $\frac{\diffd}{\diffd x}\operatorname{Ei}(-x)=e^{-x}/x$, we have
\begin{IEEEeqnarray}{lCl}
\frac{\partial}{\partial t} g(t;a) & = & \frac{e^{-\frac{1}{t}}}{(1+a t) t}\left[e^{\frac{1}{t}}a t-1-a t\right] \nonumber\\
& \geq & \frac{e^{-\frac{1}{t}}}{(1+a t) t}\left[a -1\right] \nonumber\\
& \geq & 0, \quad a\geq 1 \label{eq:appF_gincreasing}
\end{IEEEeqnarray}
where the second step follows from the lower bound $e^{\frac{1}{t}}\geq 1+\frac{1}{t}$, $t\geq 0$.

By \eqref{eq:EPvsV}, we have that $\tilde{V}_{\rho}(\xi)/\tilde{\Phi}_{\rho}(\xi)\geq 1$. It thus follows from \eqref{eq:appF_1}--\eqref{eq:appF_gincreasing} that
\begin{align}
\Sigma(\rho,\xi) & \leq g\left(\rho\tilde{\Phi}_{\rho}(\xi);\frac{\tilde{V}_{\rho}(\xi)}{\tilde{\Phi}_{\rho}(\xi)}\right) \nonumber\\
& \leq \lim_{t\to\infty} g\left(t;\frac{\tilde{V}_{\rho}(\xi)}{\tilde{\Phi}_{\rho}(\xi)}\right) \nonumber\\
& = \gamma + \log\left(\frac{\tilde{V}_{\rho}(\xi)}{\tilde{\Phi}_{\rho}(\xi)}\right), \quad \bigl(\rho>0,\,\xi\in\mathbb{C}\bigr).\label{eq:appF_almost}
\end{align} 
Maximizing the RHS of \eqref{eq:appF_almost} over $\xi\in\mathbb{C}$, and computing the limit as $\rho$ tends to infinity, gives
\begin{IEEEeqnarray}{lCl}
\varlimsup_{\rho\to\infty} \sup_{\xi\in\mathbb{C}} \Sigma(\rho,\xi) & \leq & \gamma + \varlimsup_{\rho\to\infty} \log\left(\sup_{\xi\in\mathbb{C}}\frac{\tilde{V}_{\rho}(\xi)}{\tilde{\Phi}_{\rho}(\xi)}\right) \nonumber\\
& \leq & \gamma + \log(M)
\end{IEEEeqnarray}
where the last step follows from the continuity and monotonicity of $x\mapsto \log(x)$ and from \eqref{technical_condition}. This proves Lemma~\ref{lem:g}.

\subsection{Proof of Lemma~\ref{lem:estimation_convergence}}   \label{app:lem:proof:estimation_convergence}

By the law of total probability, we have
\begin{align} 
	\Prob\bigl\{ |H| > 2\xi_0 \bigr\}
	&= \Prob\bigl\{ |H| > 2\xi_0 , |\hat{H}_{\rho}| \leq \xi_0 \bigr\} + \Prob\bigl\{ |H| > 2\xi_0 , |\hat{H}_{\rho}| > \xi_0 \bigr\} \nonumber\\
	&\leq \Prob\bigl\{ |H-\hat{H}_{\rho}| > \xi_0 \bigr\} + \Prob\bigl\{ |\hat{H}_{\rho}| > \xi_0 \bigr\} \label{total_probability}
\end{align}
using the fact that $|H| > 2\xi_0$ and $|\hat{H}_{\rho}| \leq \xi_0$ together imply that $|H-\hat{H}_{\rho}| > \xi_0$ due to the triangle inequality, and that $|H| > 2\xi_0$ and $|\hat{H}_{\rho}| > \xi_0$ together imply $|\hat{H}_{\rho}| > \xi_0$.
Using Chebyshev's inequality \cite[(4.10.7),~p.~192]{AsDo00}, the first term on the RHS of \eqref{total_probability} can be further upper-bounded by
\begin{align}   \label{V_tilde_deviation}
	\Prob\bigl\{ |H-\hat{H}_{\rho}| > \xi_0 \bigr\}
		&\leq \frac{\Exp\bigl[\tilde{V}_{\rho}(\hat{H}_{\rho})\bigr]}{\xi_0^2}.
\end{align}
Combining \eqref{V_tilde_deviation} with \eqref{total_probability} gives
\begin{equation}
\label{V_tilde_deviation_2}
\Prob\bigl\{ |H| > 2\xi_0 \bigr\} \leq \frac{\Exp\bigl[\tilde{V}_{\rho}(\hat{H}_{\rho})\bigr]}{\xi_0^2} + \Prob\bigl\{ |\hat{H}_{\rho}| > \xi_0 \bigr\}.
\end{equation}
By \eqref{eq:thm_Vtilde}, taking the limit inferior for $\rho \rightarrow \infty$ on either side of \eqref{V_tilde_deviation_2} yields
\begin{align}
	\Prob\bigl\{ |H| > 2\xi_0 \bigr\}
	&\leq \varliminf_{\rho \rightarrow \infty} \Prob\bigl\{ |\hat{H}_{\rho}| > \xi_0 \bigr\}. \label{eq:app_lem10}
\end{align}
Furthermore, the assumption that $\Prob\bigl\{H=0\bigr\}=0$, we have
\begin{equation}
	\lim_{\xi_0\downarrow 0} \Prob\bigl\{ |H| > 2\xi_0\bigr\} = \Prob\bigl\{|H|>0\bigr\} = 1.
\end{equation}
Lemma~\ref{lem:estimation_convergence} follows therefore by taking limits as $\xi_0 \downarrow 0$ on both sides of \eqref{eq:app_lem10}.

\subsection{Proof of Lemma~\ref{lem:continuity_at_0}}   \label{app:lem:proof:continuity_at_0}

For every family of random variables $\Upsilon_{\rho}$ parametrized by $\rho>0$ and taking values on $(0,\eta)$ with $\eta \in (0,\infty]$, we have by Chebyshev's inequality
\begin{equation}
\Prob\left\{\Upsilon_{\rho}>\nu\right\} \leq \frac{\Exp[\Upsilon_{\rho}]}{\nu}, \quad \textnormal{for every $\nu \in (0,\eta)$.}
\end{equation}
Using that $\lim_{\rho\to\infty} \Exp[\Upsilon_{\rho}]=0$, we thus have
\begin{equation}
\label{eq:app_cont_1}
\lim_{\rho\to\infty} \Prob\left\{\Upsilon_{\rho}>\nu\right\} = 0, \quad \textnormal{for every $\nu \in (0,\eta)$}
\end{equation}
or equivalently, $\lim_{\rho\to\infty} \Prob\left\{\Upsilon_{\rho}\leq\nu\right\} = 1$.
We upper-bound $\Exp[f(\Upsilon_{\rho})]$ for any $\nu\in(0,\eta)$ as
\begin{subequations}
\begin{IEEEeqnarray}{lCl}
\Exp[f(\Upsilon_{\rho})] & = & \Exp\left[f(\Upsilon_{\rho}) \,\operatorname{I}\{\Upsilon_{\rho}\leq\nu\}\right] + \Exp\left[f(\Upsilon_{\rho}) \,\operatorname{I}\{\Upsilon_{\rho}>\nu\}\right] \nonumber\\
& \leq & \sup_{0<t\leq\nu} f(t) \Prob\{\Upsilon_{\rho}\leq\nu\} + \sup_{\nu<t<\eta} f(t) \Prob\{\Upsilon_{\rho}>\nu\} .  \label{eq:app_cont_2}
\end{IEEEeqnarray}
Similarly, we lower-bound $\Exp[f(\Upsilon_{\rho})]$ for any $\nu\in(0,\eta)$ as
\begin{IEEEeqnarray}{lCl}
\Exp[f(\Upsilon_{\rho})] & \geq & \inf_{0<t\leq\nu} f(t) \Prob\{\Upsilon_{\rho}\leq\nu\} + \inf_{\nu<t<\eta} f(t) \Prob\{\Upsilon_{\rho}>\nu\}.   \label{eq:app_cont_3}
\end{IEEEeqnarray}
\end{subequations}
Since $f(\cdot)$ is bounded, and by \eqref{eq:app_cont_1}, taking limits for $\rho \rightarrow \infty$ in \eqref{eq:app_cont_2} and \eqref{eq:app_cont_3} gives
\begin{IEEEeqnarray}{rClClCl}
\inf_{0<t\leq\nu} f(t)
& \leq & \varliminf_{\rho \rightarrow \infty} \Exp[f(\Upsilon_{\rho})]
& \leq & \varlimsup_{\rho \rightarrow \infty} \Exp[f(\Upsilon_{\rho})]
& \leq & \sup_{0<t\leq\nu} f(t).
\end{IEEEeqnarray}
Taking the limit as $\nu$ tends to zero from above and using the continuity of $f$, we finally obtain
\begin{equation}
\lim_{\rho \rightarrow \infty} \Exp[f(\Upsilon_{\rho})] = \lim_{t \downarrow 0} f(t) = f_0
\end{equation}
which proves Lemma~\ref{lem:continuity_at_0}.

\subsection{Proof of \eqref{eq:doby_bounded_log2}}
\label{sub:doby_bounded_log2}
To prove \eqref{eq:doby_bounded_log2}, we first note that the function $x\mapsto \log^2(1+x)$ is concave for $x\geq e-1$. We thus have for an arbitrary $\delta\geq e-1$ and for $\rho\geq 1$
\begin{IEEEeqnarray}{lCl}
\IEEEeqnarraymulticol{3}{l}{\Exp\left[\log^2\left(1+\rho^{-1}+|\hat{H}_{\rho}|^2\right)\right]} \nonumber\\
\quad & = &  \Exp\left[\log^2\left(1+\rho^{-1}+|\hat{H}_{\rho}|^2\right)\operatorname{I}\left\{|\hat{H}_{\rho}|^2\leq \delta\right\}\right] + \Exp\left[\log^2\left(1+\rho^{-1}+|\hat{H}_{\rho}|^2\right)\operatorname{I}\left\{|\hat{H}_{\rho}|^2 > \delta\right\}\right] \nonumber\\
& \leq & \log^2\left(2+\delta\right)+ \Exp\left[\log^2\left(2+|\hat{H}_{\rho}|^2\right)\operatorname{I}\left\{|\hat{H}_{\rho}|^2 > \delta\right\}\right] \nonumber\\
& \leq & \log^2\left(2+\delta\right) + \Prob\left(|\hat{H}_{\rho}|^2 > \delta\right)\log^2\left(2+\frac{\Exp\left[|\hat{H}_{\rho}|^2\operatorname{I}\left\{|\hat{H}_{\rho}|^2 > \delta\right\}\right]}{\Prob\left(|\hat{H}_{\rho}|^2 > \delta\right)}\right) \IEEEeqnarraynumspace
\end{IEEEeqnarray}
where we define $0\log^2(1+a/0)\triangleq 0$ for every $a\geq 0$. Here the first inequality follows by upper-bounding $\rho^{-1}\leq 1$ in both expected values and by upper-bounding $|\hat{H}_{\rho}|^2\leq\delta$ in the first expected value, and the second inequality follows by upper-bounding the second expected value using Jensen's inequality.

We next use \eqref{eq:doby_UB_I} and \eqref{doby_highSNR_normalized} to upper-bound 
\begin{equation}
\Exp\left[|\hat{H}_{\rho}|^2\operatorname{I}\left\{|\hat{H}_{\rho}|^2 > \delta\right\}\right] \leq \Exp\left[|\hat{H}_{\rho}|^2\right] \leq 1.
\end{equation}
This yields
\begin{IEEEeqnarray}{lCl}
\Exp\left[\log^2\left(1+\rho^{-1}+|\hat{H}_{\rho}|^2\right)\right] & \leq & \log^2\left(2+\delta\right) + \Prob\left(|\hat{H}_{\rho}|^2 > \delta\right)\log^2\left(2+\frac{1}{\Prob\left(|\hat{H}_{\rho}|^2 > \delta\right)}\right) \nonumber\\
& \leq & \log^2\left(2+\delta\right) + \sup_{0<x\leq 1}\left\{x\log^2\left(2+\frac{1}{x}\right)\right\}, \quad \rho\geq 1 \label{eq:doby_(170)_1}
\end{IEEEeqnarray}
where the second step follows by maximising the second term over $\Prob\left(|\hat{H}_{\rho}|^2>\delta\right)$. Note that the supremum on the RHS of \eqref{eq:doby_(170)_1} is finite since the function $x\mapsto x\log^2(2+1/x)$ is continuous on $0<x\leq 1$ and tends to zero as $x$ tends to zero. Consequently, we have
\begin{equation}
\varlimsup_{\rho\to\infty} \Exp\left[\log^2\left(1+\rho^{-1}+|\hat{H}_{\rho}|^2\right)\right] \leq \log^2\left(2+\delta\right) + \sup_{0<x\leq 1}\left\{x\log^2\left(2+\frac{1}{x}\right)\right\} < \infty
\end{equation}
which proves \eqref{eq:doby_bounded_log2}.

\section*{Acknowledgment}
The authors wish to thank A.~Guill\'en i F\`abregas, A.~Lapidoth, A.~Martinez, and J.~Scarlett for enlightening discussions. They further wish to thank the Associate Editor A.~Tulino and the anonymous referees for their valuable comments.

\bibliographystyle{IEEEtran}
\bibliography{IEEEabrv,custom/references_t}

\begin{thebibliography}{10}
\providecommand{\url}[1]{#1}
\csname url@samestyle\endcsname
\providecommand{\newblock}{\relax}
\providecommand{\bibinfo}[2]{#2}
\providecommand{\BIBentrySTDinterwordspacing}{\spaceskip=0pt\relax}
\providecommand{\BIBentryALTinterwordstretchfactor}{4}
\providecommand{\BIBentryALTinterwordspacing}{\spaceskip=\fontdimen2\font plus
\BIBentryALTinterwordstretchfactor\fontdimen3\font minus
  \fontdimen4\font\relax}
\providecommand{\BIBforeignlanguage}[2]{{%
\expandafter\ifx\csname l@#1\endcsname\relax
\typeout{** WARNING: IEEEtran.bst: No hyphenation pattern has been}%
\typeout{** loaded for the language `#1'. Using the pattern for}%
\typeout{** the default language instead.}%
\else
\language=\csname l@#1\endcsname
\fi
#2}}
\providecommand{\BIBdecl}{\relax}
\BIBdecl

\bibitem{BiProSh98}
E.~Biglieri, J.~Proakis, and S.~Shamai~(Shitz), ``Fading channels:
  Information-theoretic and communications aspects,'' \emph{IEEE Transactions
  on Information Theory}, vol.~44, pp. 2619--2692, 1998.

\bibitem{Me00}
M.~M\'edard, ``The effect upon channel capacity in wireless communications of
  perfect and imperfect knowledge of the channel,'' \emph{IEEE Transactions on
  Information Theory}, vol.~46, no.~3, pp. 933--946, May 2000.

\bibitem{coverthomas91}
T.~M. Cover and J.~A. Thomas, \emph{Elements of Information Theory},
  1st~ed.\hskip 1em plus 0.5em minus 0.4em\relax John Wiley \& Sons, 1991.

\bibitem{LaMo03}
A.~Lapidoth and S.~M. Moser, ``Capacity bounds via duality with applications to
  multiple-antenna systems on flat-fading channels,'' \emph{IEEE Transactions
  on Information Theory}, vol.~49, no.~10, pp. 2426--2467, Oct. 2003.

\bibitem{LaSh02}
A.~Lapidoth and S.~Shamai~(Shitz), ``Fading channels: How perfect need `perfect
  side information' be?'' \emph{IEEE Transactions on Information Theory},
  vol.~48, no.~5, pp. 1118--1134, May 2002.

\bibitem{DeCo91}
A.~Dembo and T.~Cover, ``Information theoretic inequalities,'' \emph{IEEE
  Transactions on Information Theory}, vol.~37, no.~6, pp. 1501--1518, Nov.
  1991.

\bibitem{BaFoMe01}
J.~Baltersee, G.~Fock, and H.~Meyr, ``Achievable rate of {MIMO} channels with
  data-aided channel estimation and perfect interleaving,'' \emph{IEEE Journal
  on Selected Areas in Communications}, vol.~19, no.~12, pp. 2358--2368, Dec.
  2001.

\bibitem{YoGo06}
T.~Yoo and A.~Goldsmith, ``Capacity and power allocation for fading {MIMO}
  channels with channel estimation error,'' \emph{IEEE Transactions on
  Information Theory}, vol.~52, no.~5, pp. 2203--2214, May 2006.

\bibitem{GrSz84}
U.~Grenander and G.~Szeg\H{o}, \emph{Toeplitz Forms and their Applications},
  2nd~ed.\hskip 1em plus 0.5em minus 0.4em\relax Chelsea Publishing Company,
  1984.

\bibitem{La05}
A.~Lapidoth, ``On the asymptotic capacity of stationary {G}aussian fading
  channels,'' \emph{IEEE Transactions on Information Theory}, vol.~51, no.~2,
  pp. 437--446, Feb. 2005.

\bibitem{DoToSa04}
M.~Dong, L.~Tong, and B.~M. Sadler, ``Optimal insertion of pilot symbols for
  transmissions over time-varying flat-fading channels,'' \emph{IEEE
  Transactions on Signal Processing}, vol.~21, no.~5, pp. 1403--1418, May 2004.

\bibitem{Lo08}
A.~Lozano, ``Interplay of spectral efficiency, power and {D}oppler spectrum for
  reference-signal-assisted wireless communication,'' \emph{IEEE Transactions
  on Wireless Communications}, vol.~7, no.~12, pp. 5020--5029, Dec. 2008.

\bibitem{AsKoGu11}
A.~Asyhari, T.~Koch, and A.~Guill\'en~i F\`abregas, ``Nearest neighbour
  decoding and pilot-aided channel estimation in stationary {G}aussian
  flat-fading channels,'' in \emph{Proc. IEEE International Symposium on
  Information Theory}, August 2011, pp. 2786--2790.

\bibitem{AsKoGu13}
\BIBentryALTinterwordspacing
------, ``Nearest neighbor decoding and pilot-aided channel estimation for
  fading channels,'' January 2013. [Online]. Available:
  \url{http://arxiv.org/abs/1301.1223}
\BIBentrySTDinterwordspacing

\bibitem{OhGi02}
S.~Ohno and G.~Giannakis, ``Average-rate optimal {PSAM} transmissions over
  time-selective fading channels,'' \emph{IEEE Transactions on Wireless
  Communications}, vol.~1, no.~4, pp. 712--720, Oct. 2002.

\bibitem{KaSh93}
G.~Kaplan and S.~Shamai~(Shitz), ``Information rates of compound channels with
  application to antipodal signaling in a fading environment,'' \emph{AEU},
  vol.~47, no.~4, pp. 228--239, 1993.

\bibitem{Sti66}
I.~G. Stiglitz, ``Coding for a class of unknown channels,'' \emph{IEEE
  Transactions on Information Theory}, vol.~12, no.~2, pp. 189--195, Apr. 1966.

\bibitem{GaLaTe00}
A.~Ganti, A.~Lapidoth, and I.~E. Telatar, ``Mismatched decoding revisited:
  General alphabets, channels with memory, and wide-band limit,'' \emph{IEEE
  Transactions on Information Theory}, vol.~46, no.~7, pp. 2315--2328, Nov.
  2000.

\bibitem{La96}
A.~Lapidoth, ``Mismatched decoding and the multiple-access channel,''
  \emph{IEEE Transactions on Information Theory}, vol.~42, no.~5, pp.
  1439--1452, Sept. 1996.

\bibitem{PaKoFo12}
A.~Pastore, T.~Koch, and J.~Fonollosa, ``Improved capacity lower bounds for
  fading channels with imperfect {CSI} using rate splitting,'' in
  \emph{Proceedings IEEE 27th Convention of Electrical Electronics Engineers in
  Israel (IEEEI)}, Nov. 2012, pp. 1--5.

\bibitem{AsDo00}
R.~B. Ash and C.~A. Dol\'eans-Dade, \emph{Probability and Measure Theory},
  2nd~ed.\hskip 1em plus 0.5em minus 0.4em\relax Elsevier/Academic Press, 2000.

\bibitem{GrRy94}
I.~S. Gradshteyn and I.~M. Ryzhik, \emph{Table of Integrals, Series, and
  Products}, 6th~ed.\hskip 1em plus 0.5em minus 0.4em\relax Academic Press,
  2000.

\end{thebibliography}

\end{document}